\documentclass[aps,prd,nofootinbib,showkeys,preprint,
floatfix]{revtex4-1}

\usepackage{graphicx}
\usepackage{graphics}
\usepackage[mathscr]{eucal}
\usepackage{amssymb}
\usepackage{amsmath}
\usepackage{xspace}
\usepackage{listings}
\usepackage{ulem}
\usepackage{color} 

\definecolor{brown}{rgb}{0.8,0.4,0}

\newcommand{\DRbar}[0]{{\ensuremath{\overline{\mathrm{DR}}}}}
\newcommand{\SARAH}[0]{{\tt SARAH}\xspace}
\newcommand{\SPheno}[0]{{\tt SPheno}\xspace}

\newcommand{\sign}[0]{\text{sign}\xspace}

\newcommand{\BL}[0]{{\ensuremath{B-L}}\xspace}
\newcommand{\UBL}[0]{{\ensuremath{U(1)_{B-L}}}\xspace}

\newcommand{\gBL}[1]{{\ensuremath{g_{BL}^{#1}}}\xspace}
\newcommand{\gmix}{{\ensuremath{\bar{g}}}\xspace}
\newcommand{\mReSnuSq}[0]{{\ensuremath{m^{2}_{{\tilde{\nu}}^{S}}}}\xspace}
\newcommand{\mImSnuSq}[0]{{\ensuremath{m^{2}_{{\tilde{\nu}}^{P}}}}\xspace}

\newcommand{\vevs}[0]{\textit{vev}s\xspace}
\newcommand{\msugra}[0]{mSUGRA\xspace}

\newcommand{\EQ}[0]{Eq.\xspace}
\newcommand{\EQS}[0]{Eqs.\xspace}
\newcommand{\TAB}[0]{Table\xspace}
\newcommand{\FIG}[0]{Fig.\xspace}
\newcommand{\FIGS}[0]{Figs.\xspace}
\newcommand{\REF}[1]{Ref.~\cite{#1}}

\newcommand{\EG}{\textit{e.g}.\xspace}
\newcommand{\IE}{\textit{i.e}.\xspace}
\def\gsim{\raise0.3ex\hbox{$\;>$\kern-0.75em\raise-1.1ex\hbox{$\sim\;$}}}
\def\lsim{\raise0.3ex\hbox{$\;<$\kern-0.75em\raise-1.1ex\hbox{$\sim\;$}}}

\newcommand{\emiss}{E\!\!\!/}
\newcommand{\zp}{\ensuremath{Z'}}

\newcommand{\mzp}{\ensuremath{M_{Z'}}}

\newcommand{\AddrBonn}{%
Bethe Center for Theoretical Physics \& Physikalisches Institut der
 Universit\"at Bonn, \\
53115 Bonn, Germany }

\newcommand{\AddrWur}{%
Institut f\"ur Theoretische Physik und Astronomie,
Universit\"at W\"urzburg\\
Am Hubland,
97074 W\"urzburg, Germany}

\begin{document}

\title{Implications of gauge kinetic mixing on
$Z'$ and slepton production at the LHC}

\author{M.\ E.\ Krauss} \email{manuel.krauss@physik.uni-wuerzburg.de}

\author{B.\ O'Leary} \email{ben.oleary@physik.uni-wuerzburg.de}

\author{W.\ Porod} \email{porod@physik.uni-wuerzburg.de}\affiliation{\AddrWur}

\author{F.\ Staub}\email{fnstaub@th.physik.uni-bonn.de}
\affiliation{\AddrBonn}



\preprint{Bonn-TH-2012-10}

\begin{abstract}
We consider a supersymmetric version of the standard model extended
by an additional $U(1)_{B-L}$. This model can be embedded in
an mSUGRA-inspired model where the mass parameters of the scalars
and gauginos unify at the scale of grand unification.
In this class of models the renormalization group equation evolution of gauge couplings as well
as of the soft SUSY-breaking parameters require the proper treatment 
of gauge kinetic mixing. We first show that this has a profound
impact on the phenomenolgy of the $Z'$ and as a consequence
the current LHC bounds on its mass are reduced significantly
from about 1970 GeV to 1790 GeV.
They are even further reduced if the $Z'$ can decay into
supersymmetric particles. Secondly, we show that in this way sleptons
can be produced at the LHC in the 14 TeV phase
with masses of several hundred GeV. In the case 
of squark and gluino masses in the multi-TeV range, this might
 become
an important discovery channel for sleptons up to  800 GeV (900 GeV)
for an integrated luminosity of 100 fb$^{-1}$ (300 fb$^{-1}$).
\end{abstract}

\maketitle

\section{Introduction}

The LHC is rapidly extending our knowledge of
 the TeV scale. As there are currently no signs of new physics, models
beyond the standard model (SM) have begun to be severely constrained.
One of the most popular model classes
 is that of supersymmetric extensions,
in particular the minimal supersymmetric standard model (MSSM).
As the MSSM itself has over 100 free parameters, mainly models
with a smaller set of parameters, like the constrained MSSM (CMSSM)
 with five parameters \cite{Nilles:1983ge,Drees:2004jm},
are studied. This model
has been fitted to various combinations of experimental data: see
\EG\cite{Allanach:2011ut,Buchmueller:2011aa,Allanach:2011wi,Bertone:2011nj%
,Buchmueller:2011ki,Buchmueller:2011sw,Beskidt:2012bh,Roszkowski:2012uf%
,Bechtle:2012zk}, indicating that gluinos and squarks have masses
 in
the multi-TeV range. In particular, if one wants to explain
the recent hints of a SM-like Higgs boson with mass of 125~GeV
\cite{ATLAS:2012ae,Chatrchyan:2012tx},
the CMSSM becomes more and more unlikely \cite{Bechtle:2012zk}.
However,  this is mainly due to the strong correlations between
the various masses of the supersymmetric particles. Once these
are given up, a light SUSY spectrum is still compatible with LHC
data as discussed in 
\cite{Berger:2008cq,Conley:2010du,Sekmen:2011cz,LeCompte:2011fh,%
Arbey:2011un}.

In the framework of constrained models, a way out of this tension
is to consider extended models such as the next-to-minimal supersymmetric standard model (NMSSM)
 \cite{Maniatis:2009re,Ellwanger:2009dp}. Indeed,
it has been shown that even in the
 constrained NMSSM a Higgs mass of 125~GeV 
can be explained \cite{Ellwanger:2011aa,Gunion:2012zd} 
and in the generalized NMSSM these masses are obtained with even less
fine-tuning \cite{Ross:2012nr}.
A second possibility is to consider models with extended gauge
structures, since then the upper bound on the
 lightest Higgs boson is also relaxed due to additional tree-level
 contributions
\cite{Haber:1986gz,Drees:1987tp,Cvetic:1997ky,Nie:2001ti,%
Ma:2011ea,Hirsch:2011hg}. Such models arise naturally in the
context of embedding the SM gauge group in a larger group such
as $SO(10)$ or $E_6$: see \EG
 \cite{Cvetic:1983su,Cvetic:1997wu,Aulakh:1997ba,Aulakh:1997fq,
FileviezPerez:2008sx,Siringo:2003hh}, which can also
nicely explain neutrino data via the seesaw mechanism
\cite{Minkowski:1977sc,Yanagida:1979as,Mohapatra:1979ia,Schechter:1980gr}. 
It can be shown that such
models can have a $Z'$ with a mass in the TeV range 
\cite{Malinsky:2005bi,DeRomeri:2011ie}. $Z'$ searches are
therefore among the targets of the Tevatron and LHC collaborations and
bounds on its mass have been set 
\cite{Aaltonen:2011gp,Abazov:2010ti,Collaboration:2011dca,%
Chatrchyan:2011wq}. These bounds depend on the concrete gauge group
and the couplings of the $Z'$ to the SM fermions: see \EG
\cite{Salvioni:2009mt,Salvioni:2009jp,Basso:2010pe,Jezo:2012rm,Cao:2012ng}.
 For reviews on various $Z'$ models
see \EG \cite{Leike:1998wr,Langacker:2008yv}.
 
It has been known for some time that decays into supersymmetric
particles can also strongly impact the phenomenology of the $Z'$
\cite{Gherghetta:1996yr,Erler:2002pr}. The LHC phenomenology
of supersymmetric $U(1)$ extensions has been discussed
in the context of $SO(10)$ and $E_6$ embeddings in
\cite{Kang:2004bz,Chang:2011be,Corcella:2012dw} 
and in a more general class of models with a $U(1)_{B-x L}$
 in  \cite{Baumgart:2006pa}. 
In \REF{Chang:2011be}  a $U(1)_{B-L}$ extension of the
MSSM is also discussed.
Furthermore, it has  been 
pointed out that 
sleptons, charginos, and neutralinos can be directly produced
via a heavy $Z'$
at the LHC even if they have masses of hundreds of GeV.

$U(1)$ extensions of the SM have a peculiar feature, namely,
gauge kinetic mixing \cite{Holdom:1985ag,Babu:1997st,delAguila:1988jz}.
In the previous studies either it has been
argued that these effects are small or they have been completely
ignored. However, it has also been pointed out in 
 \cite{Rizzo:1998ut,Rizzo:2012rf} that gauge kinetic mixing can
 be important in $Z'$ searches in the case of $E_6$ embeddings.
Moreover, it has recently been shown that gauge kinetic
mixing is important for the spectrum in $U(1)_{B-L}$ 
extensions of the MSSM \cite{O'Leary:2011yq}. 
 Such models can be embedded, for example,
in a string-inspired
$E_8 \times E_8$ gauge group
\cite{Buchmuller:2006ik,Ambroso:2009sc,Ambroso:2010pe}.

In this paper we will explicitly show how existing collider
bounds on the $Z'$ mass are
changed once gauge kinetic mixing is taken into account. We find
that this can reduce the bounds by about 200~GeV, as the couplings of
$Z'$ to the SM fermions depend on this mixing. This clearly
 affects
also the cross section for SUSY particles produced via a $Z'$.
However, as we will demonstrate, LHC with $\sqrt{s}= 14$ TeV and
a luminosity of 100 fb$^{-1}$ should be able to discover
sleptons via a $Z'$ with slepton  masses up to 800~GeV,
provided the $Z'$ is not much heavier than 2.8~TeV. In the case of 
300 fb$^{-1}$, this extends to slepton masses of about 900~GeV up
to $Z'$ masses of 3.1~TeV.

The remainder of this paper is organized
as follows: In Sec. \ref{sec:model} we briefly summarize the
main features of the model and its particle content. In 
Sec. \ref{sec:results} we first discuss how gauge kinetic mixing
as well as supersymmetric final states  affect the $Z'$ phenomenology 
in the context of
constrained models. Afterwards we discuss the possibilities of the
LHC to discover SUSY particles, in particular charged sleptons,
 via a $Z'$.
Here we will depart from the universality assumption as this mainly
depends on the slepton and $Z'$ masses. Finally we draw our conclusions
in Sec. \ref{sec:conclusions}. In the Appendix we give the
couplings of the $Z'$ to the scalars and fermions, including
terms arising due to gauge kinetic mixing effects.

\section{The Model}
\label{sec:model}

In this section we discuss briefly the particle content and the
superpotential of the model under consideration and we give the
tree-level masses and mixings of the particles important to
 our studies. For a detailed discussion of the masses of
all particles as well as of the corresponding one-loop corrections, we
refer to \cite{O'Leary:2011yq}. In addition, we recall the main aspects
of $U(1)$ kinetic mixing since it has important consequences for
the phenomenology of the $Z'$ within this model.

\subsection{Particle content and superpotential}
The model consists of three generations of matter particles including
right-handed neutrinos which can, for example, be embedded in $SO(10)$
16-plets
\cite{Buchmuller:2006ik,Ambroso:2009sc,Ambroso:2010pe}. 
Moreover, below the grand unified theory (GUT) scale the usual MSSM Higgs doublets
are present, as well as two fields $\eta$ and $\bar{\eta}$ responsible
for the breaking of the \UBL. The vacuum expectation value of $\eta$ 
induces a  Majorana mass term for the right-handed neutrinos. 
Thus we interpret the \BL charge of this field as its lepton number,
and likewise for $\bar{\eta}$, and call these fields bileptons since
they carry twice the usual lepton number.  A summary
of the quantum numbers of the chiral superfields with respect to 
$SU(3)_C \times SU(2)_L \times U(1)_Y \times  \UBL$ is given 
in \TAB~\ref{tab:cSF}.
\begin{table} 
\centering
\begin{tabular}{|c|c|c|c|c|c|} 
\hline \hline 
Superfield & Spin 0 & Spin \(\frac{1}{2}\) & Generations &
Quantum numbers  \\ 
\hline 
\(\hat{Q}\) & \(\tilde{Q}\) & \(Q\) & 3
 & \(({\bf 3},{\bf 2},\frac{1}{6},\frac{1}{6}) \) \\ 
\({\hat{d}}^{c}\) & \(\tilde{d}^c\) & \(d^c\) & 3
 & \(({\bf \overline{3}},{\bf 1},\frac{1}{3},-\frac{1}{6}) \) \\ 
\({\hat{u}}^{c}\) & \(\tilde{u}^c\) & \(u^c\) & 3
 & \(({\bf \overline{3}},{\bf 1},-\frac{2}{3},-\frac{1}{6}) \) \\ 
\(\hat{L}\) & \(\tilde{L}\) & \(L\) & 3
 & \(({\bf 1},{\bf 2},-\frac{1}{2},-\frac{1}{2}) \) \\ 
\({\hat{e}}^{c}\) & \(\tilde{e}^c\) & \(e^c\) & 3
 & \(({\bf 1},{\bf 1},1,\frac{1}{2}) \) \\ 
\({\hat{\nu}^{c}}\) & \(\tilde{\nu}^c\) & \(\nu^c\) & 3
 & \(({\bf 1},{\bf 1},0,\frac{1}{2}) \) \\ 
\(\hat{H}_d\) & \(H_d\) & \(\tilde{H}_d\) & 1
 & \(({\bf 1},{\bf 2},-\frac{1}{2},0) \) \\ 
\(\hat{H}_u\) & \(H_u\) & \(\tilde{H}_u\) & 1
 & \(({\bf 1},{\bf 2},\frac{1}{2},0) \) \\ 
\(\hat{\eta}\) & \(\eta\) & \(\tilde{\eta}\) & 1
 & \(({\bf 1},{\bf 1},0,-1) \) \\ 
\(\hat{\bar{\eta}}\) & \(\bar{\eta}\) & \(\tilde{\bar{\eta}}\) & 1
 & \(({\bf 1},{\bf 1},0,1) \) \\ 
\hline \hline
\end{tabular} 
\caption{Chiral superfields and their quantum numbers with
respect to \( SU(3)_C\otimes\, SU(2)_L\otimes\,  U(1)_Y\otimes\,
\UBL\).}
\label{tab:cSF}
\end{table}

The superpotential is given by
\begin{align} 
\nonumber 
W = & \, Y^{ij}_u\,{\hat{u}}^{c}_i\,\hat{Q}_j\,\hat{H}_u\,
- Y_d^{ij} \,{\hat{d}}^{c}_i\,\hat{Q}_j\,\hat{H}_d\,
- Y^{ij}_e \,{\hat{e}}^{c}_i\,\hat{L}_j\,\hat{H}_d\,
+\mu\,\hat{H}_u\,\hat{H}_d\, \\
 & \, \, 
+Y^{ij}_{\nu}\,{\hat{\nu}}^{c}_i\,\hat{L}_j\,\hat{H}_u\,
- \mu' \,\hat{\eta}\,\hat{\bar{\eta}}\,
+Y^{ij}_x\,{\hat{\nu}}^{c}_i\,\hat{\eta}\,{\hat{\nu}}^{c}_j\, ,
\label{eq:superpot}
\end{align} 
and we have the additional soft SUSY-breaking terms:
\begin{align}
\nonumber \mathscr{L}_{SB} = & \mathscr{L}_{MSSM}
 - \lambda_{\tilde{B}} \lambda_{\tilde{B}'} {M}_{B B'}
 - \frac{1}{2} \lambda_{\tilde{B}'} \lambda_{\tilde{B}'} {M}_{B'}
 - m_{\eta}^2 |\eta|^2 - m_{\bar{\eta}}^2 |\bar{\eta}|^2
 - {m_{{\nu}^{c},ij}^{2}} (\tilde{\nu}_i^c)^* \tilde{\nu}_j^c \\
& - \eta \bar{\eta} B_{\mu'} + T^{ij}_{\nu}  H_u \tilde{\nu}_i^c \tilde{L}_j
 + T^{ij}_{x} \eta \tilde{\nu}_i^c \tilde{\nu}_j^c  \, ,
\end{align}
where $i,j$ are generation indices. Without loss of generality one can
 take $B_\mu$
and $B_{\mu'}$ to be real.
 The extended gauge group breaks to
$SU(3)_C \otimes U(1)_{em}$ as the Higgs fields and bileptons receive vacuum
expectation values (\vevs):
\begin{align} 
H_d^0 = & \, \frac{1}{\sqrt{2}} \left(\sigma_{d} + v_d  + i \phi_{d} \right),
\hspace{1cm}
H_u^0 = \, \frac{1}{\sqrt{2}} \left(\sigma_{u} + v_u  + i \phi_{u} \right)\\ 
\eta
= & \, \frac{1}{\sqrt{2}} \left(\sigma_\eta + v_{\eta} + i \phi_{\eta} \right),
\hspace{1cm}
\bar{\eta}
= \, \frac{1}{\sqrt{2}} \left(\sigma_{\bar{\eta}} + v_{\bar{\eta}}
 + i \phi_{\bar{\eta}} \right) \, .
\end{align} 
We define $\tan\beta' = ( v_{\eta} / v_{\bar{\eta}} )$ in analogy to
$\tan\beta = ( v_u / v_d )$ in the MSSM.

\subsection{Gauge kinetic mixing}
\label{subsec:kineticmixing}

As already mentioned in the Introduction, the presence of two Abelian
gauge groups in combination with the given particle content gives rise
to a new effect absent in the MSSM or other SUSY models with just one
Abelian gauge group: gauge kinetic mixing.
In the Lagrangian, the combination
\begin{equation}
\label{eq:offfieldstrength}
- \chi_{ab}  \hat{F}^{a, \mu \nu} \hat{F}^b_{\mu \nu}, \quad a \neq b
\end{equation}
of the field-strength tensors is allowed by gauge and Lorentz
invariance \cite{Holdom:1985ag} because $\hat{F}^{a, \mu \nu}$ and 
$\hat{F}^{b, \mu \nu}$ are gauge invariant quantities by themselves,

Even if such a term is absent at a given scale, it will be induced
by renormalization group equation (RGE) running. 
This can be seen most
easily by inspecting the matrix of the anomalous dimension, which at
one loop is given by
\begin{equation}
\gamma_{ab} = \frac{1}{16 \pi^2} \mbox{Tr}Q_a Q_b \,,
\end{equation}
where the indices $a$ and $b$ run over all $U(1)$ groups and
the trace runs over all fields with charge $Q$ under the corresponding
$U(1)$ group.

For our model we 
obtain in the basis $(U(1)_Y, U(1)_{B-L})$
\begin{equation}
\gamma = \frac{1}{16 \pi^2} N \left( \begin{array}{cc} 11 & 4 \\
 4 & 6 \end{array} \right) N,
\end{equation}
and we see that there are sizable off-diagonal elements. $N$ contains
the GUT normalization of the two Abelian gauge groups. We will take as
in ref.~\cite{O'Leary:2011yq} \(\sqrt{\frac{3}{5}}\) for
\(U(1)_{Y}\) and \(\sqrt{\frac{3}{2}}\) for \UBL, \IE
$N=\text{diag}(\sqrt{\frac{3}{5}},\sqrt{\frac{3}{2}})$ and
obtain 
\begin{equation}
\label{eq:gammaMatrix}
 \gamma = \frac{1}{16 \pi^2}
  \left( \begin{array}{cc} \frac{33}{5} & 6 \sqrt{\frac{2}{5}} \\
                           6 \sqrt{\frac{2}{5}} & 9 \end{array} \right) .
\end{equation}
The largeness of the off-diagonal terms indicate that 
sizable $U(1)$  kinetic mixing terms are induced via RGE evaluation at 
lower scales, even if at the GUT scale they are zero.  In
practice it turns out that it is easier to work with noncanonical
covariant derivatives instead of off-diagonal field-strength tensors
such as in \EQ~(\ref{eq:offfieldstrength}).  However, both approaches
are equivalent \cite{Fonseca:2011vn}.  Hence in the following we
consider covariant derivatives of the form
\begin{equation}
\label{eq:kovariantDerivative}
 D_\mu  = \partial_\mu - i Q_{\phi}^{T} G  A \,,
\end{equation}
where \(Q_{\phi}\) is a vector containing the charges of the field $\phi$ with
respect to the two Abelian gauge groups, $G$ is the gauge coupling matrix
\begin{equation}
 G = \left( \begin{array}{cc} g_{YY} & g_{YB} \\
                              g_{BY} & g_{BB} \end{array} \right)\, ,
\end{equation}
and $A$ contains the gauge bosons $A = ( A^Y_\mu, A^B_\mu )^T$.

As long as the two Abelian gauge groups are unbroken, the following
change of basis is always possible:
\begin{equation}
A= \left( \begin{array}{c}
A^Y_\mu \\ A^B_\mu
\end{array} \right) \to
A'= \left( \begin{array}{c}
A^Y_\mu{}' \\ A^B_\mu{}'
\end{array} \right) = R A\, ,
\end{equation}
where $R$ is an
orthogonal $2\times 2$ matrix.  
This freedom can be used to choose a basis such
that electroweak precision data can be accommodated easily. A particular
convenient choice is the basis where \(g_{B Y}=0\) because then
only the Higgs doublets contribute to the entries in the gauge boson
mass matrix of the $U(1)_Y\otimes SU(2)_L$ sector and the impact of
$\eta$ and $\bar{\eta}$ is only in the off-diagonal elements as
discussed in Sec. \ref{subsec:gaugebosons}.  Therefore, we choose
the following basis at the electroweak scale \cite{Chankowski:2006jk}:
\begin{align}
\label{eq:gYYp}
 g'_{YY}
 = & \frac{g_{YY} g_{B B} - g_{Y B} g_{B Y}}{\sqrt{g_{B B}^2 + g_{B Y}^2}}
 = g_1 \, , \\
 g'_{BB} = & \sqrt{g_{B B}^2 + g_{B Y}^2} = \gBL{}  \, , \\
 \label{eq:gtilde}
 g'_{Y B}
 = & \frac{g_{Y B} g_{B B} + g_{B Y} g_{YY}}{\sqrt{g_{B B}^2 + g_{B Y}^2}}
 = \gmix \, , \\
 g'_{B Y} = & 0\, .
\label{eq:gBYp}
\end{align}

\subsection{Tadpole equations}
\label{subsec:tadpoles}
Having in mind mSUGRA-like boundary conditions for the soft
 SUSY-breaking
parameters,
we solve the tadpole equations arising from the
minimization conditions
of the vacuum with respect to  \(|\mu|^2, B_\mu, |\mu'|^2\) and \(B_{\mu'}\).
Using $x^2=v_{\eta}^{2} + v_{\bar{\eta}}^{2}$ and $v^2=v_{d}^{2}+
v_{u}^{2}$, we obtain at tree level
\begin{align}
\label{eq:tadmu}
 |\mu|^2 = & \frac{1}{8} \Big(\Big(2 \gmix  \gBL{}  x^{2}
 \cos(2 {\beta'})
    -4 m_{H_d}^2  + 4 m_{H_u}^2 \Big)\sec(2 \beta)
    -4 \Big(m_{H_d}^2 + m_{H_u}^2\Big)
 - \Big(g_{1}^{2} + \gmix ^{2} + g_{2}^{2}\Big)v^{2} \Big)\, ,\\
 B_\mu =
 &-\frac{1}{8} \Big(-2 \gmix  \gBL{}  x^{2} \cos(2 {\beta'})
    + 4 m_{H_d}^2  -4 m_{H_u}^2
  + \Big(g_{1}^{2} + \gmix ^{2} + g_{2}^{2}\Big)v^{2} \cos(2 \beta)
   \Big)\tan(2 \beta )  \, ,    \\
 |\mu'|^2 =& \frac{1}{4} \Big(-2 \Big(\gBL{2} x^{2}
  + m_{\eta}^2 + m_{\bar{\eta}}^2\Big) + \Big(2 m_{\eta}^2  -2 m_{\bar{\eta}}^2
  + \gmix  \gBL{}  v^{2} \cos(2 \beta) \Big)\sec(2 {\beta'}) \Big)\, ,
 \\
\label{eq:tadBmuP}
 B_{\mu'} =&  \frac{1}{4} \Big(-2 \gBL{2} x^{2} \cos(2 {\beta'}) 
   + 2 m_{\eta}^2  -2 m_{\bar{\eta}}^2  + \gmix  \gBL{}  v^{2}
 \cos(2 \beta)
   \Big)\tan(2 {\beta'} )\, .
\end{align}
In the numerical evaluation we take also the one-loop
 corrections
into account as discussed in \cite{O'Leary:2011yq}. The phases of 
$\mu$ and $\mu'$ are not fixed via the tadpole  equations
 and  thus are taken as additional input parameters. 
As the phases are not important for our considerations, we set
them to zero, \EG $\text{sign}(\mu),\text{sign}(\mu')>0$. 

\subsection{Gauge boson mixing}
\label{subsec:gaugebosons}
Because of the presence of the kinetic mixing terms, the $B'$ boson mixes
at tree level with the $B$ and $W^3$ bosons. Requiring the conditions of
\EQS~(\ref{eq:gYYp})-(\ref{eq:gBYp}) means that the corresponding mass
matrix reads, in the basis $(B,W^3,B')$,
\begin{align}
\label{eq:MMgauge}
\left(\begin{array}{ccc}
\frac{1}{4} g_{1}^{2} v^2 & -\frac{1}{4} g_1 g_2 v^2
 & \frac{1}{4} g_1 \gmix  v^2\\
-\frac{1}{4} g_1 g_2 v^2 & \frac{1}{4} g_{2}^{2} v^2
 & -\frac{1}{4}\gmix  g_2 v^2\\
\frac{1}{4} g_1 \gmix  v^2 & -\frac{1}{4}\gmix  g_2 v^2
 & ( \gBL{2} x^2 + \frac{1}{4} \gmix ^{2} v^2 )
\end{array} \right) \, .
\end{align}
In the limit $\gmix  \rightarrow 0 $ both sectors decouple and the
upper $2\times 2$ block is just the standard mass matrix of the
neutral gauge bosons in electroweak symmetry breaking. This mass matrix  can
be diagonalized by a unitary mixing matrix to get the physical mass
eigenstates $\gamma$, $Z$, and $Z'$. 
Because of the special form of this matrix, the corresponding
rotation matrix can be expressed
by two mixing angles $\Theta_W$ and $\Theta'_W$ as
\begin{align}
\left(\begin{array}{c}
B\\
W\\
{B'}\end{array} \right)
 = & \,\left(
\begin{array}{ccc}
\cos\Theta_W & -\cos{\Theta'}_W \sin\Theta_W & \sin\Theta_W \sin{\Theta'}_W \\
\sin\Theta_W &  \cos\Theta_W \cos{\Theta'}_W & -\cos\Theta_W \sin{\Theta'}_W \\
 0 & \sin{\Theta'}_W  & \cos{\Theta'}_W \end{array}
\right)
\left(\begin{array}{c}
\gamma\\
Z\\
{Z'}\end{array} \right) \, ,
\end{align}
where ${\Theta'}_W$ can be approximated by \cite{Basso:2010jm}
\begin{equation}
\label{eq:ThetaWP}
\tan 2 {\Theta'}_W \simeq
 \frac{2 \gmix  \sqrt{g_1^2 + g_2^2}}{\gmix ^2
 + 16 ( x / v )^{2} \gBL{2}
 -g_2^2 - g_1^2}\, .
\end{equation}
The exact eigenvalues of \EQ~(\ref{eq:MMgauge}) are given by
\begin{align}
M_{\gamma} &= 0 \, , \\
M^2_{Z,Z'} & = \frac{1}{8}\Big((g_1^2+g_2^2+\gmix ^2)v^2
 + 4 \gBL{2} x^2 \mp \nonumber \\
 & \hspace{1.5cm} \sqrt{(g_1^2+g_2^2+\gmix ^2)^2 v^4
 - 8 (g_1^2+g_2^2-\gmix^2) \gBL{2} v^2 x^2
 + 16 \gBL{4} x^4}\Big)\, .
\end{align}
Expanding these formulas in powers of $v^2/x^2$, we find up to first
order:
\begin{equation}
\label{eq:MassZP}
M^2_Z = \frac{1}{4}\left(g_1^2 + g_2^2\right) v^2 \,,\hspace{1cm}
M^2_{Z'} = \gBL{2} x^2 + \frac{1}{4} \gmix ^2 v^2 \, .
\end{equation}
All parameters in \EQ~(\ref{eq:tadmu})-(\ref{eq:MMgauge}) 
as well as in the
 following mass matrices
 are understood as
running parameters at a given renormalization scale $\tilde Q$. 
Note that the \vevs
 $v_d$ and $v_u$
are obtained from the running mass $M_Z(\tilde Q)$ of the $Z$ boson,
which is related to the pole mass $M_Z$ through
\begin{equation}
M^2_Z(\tilde Q) = \frac{g^2_1+g^2_2}{4} (v^2_u+v^2_d)
 = M^2_Z + \mathrm{Re}\big\{ \Pi^T_{ZZ}(M^2_Z) \big\}.
\end{equation}
Here, $\Pi^T_{ZZ}$ is the transverse self-energy of the $Z$. See for more
 details also \REF{Pierce:1996zz}.

\subsection{Neutralinos}
\label{subsec:neutralinos}

In the neutralino sector, the gauge kinetic effects lead
to a mixing between the usual MSSM neutralinos with the additional
states. Both sectors would decouple  were these to be neglected.  The
mass matrix reads in the basis \( \left(\lambda_{\tilde{B}},
  \tilde{W}^0, \tilde{H}_d^0, \tilde{H}_u^0, \lambda_{\tilde{B}{}'},
  \tilde{\eta}, \tilde{\bar{\eta}}\right) \)
\begin{equation} 
\label{eq:NeutralinoMM}
m_{\tilde{\chi}^0} = \left( 
\begin{array}{ccccccc}
M_1 & 0 & -\frac{1}{2} g_1 v_d & \frac{1}{2} g_1 v_u & \frac{1}{2} {M}_{B B'}
 & 0 & 0 \\ 
0 & M_2 & \frac{1}{2} g_2 v_d  & -\frac{1}{2} g_2 v_u  & 0 & 0 & 0 \\ 
-\frac{1}{2} g_1 v_d  & \frac{1}{2} g_2 v_d  & 0 & - \mu
 & -\frac{1}{2} \gmix  v_d  & 0 & 0 \\ 
\frac{1}{2} g_1 v_u  & -\frac{1}{2} g_2 v_u & - \mu  & 0
 & \frac{1}{2} \gmix  v_u  & 0 & 0 \\ 
\frac{1}{2} {M}_{B B'}  & 0 & -\frac{1}{2} \gmix  v_d
 & \frac{1}{2} \gmix  v_u  & {M}_{B} & - \gBL{}  v_{\eta}
 & \gBL{}  v_{\bar{\eta}} \\ 
0 & 0 & 0 & 0 & - \gBL{}  v_{\eta}  & 0 & - {\mu'} \\ 
0 & 0 & 0 & 0 & \gBL{}  v_{\bar{\eta}}  & - {\mu'} & 0\end{array} 
\right) \, .
\end{equation} 
It is well known that for real parameters such a matrix can be
diagonalized by an orthogonal mixing matrix $N$ such that $N^*
M^{\tilde\chi^0}_T N^\dagger$ is diagonal. For complex parameters one
has to diagonalize $M^{\tilde\chi^0}_T (M^{\tilde\chi^0}_T)^\dagger$.

\subsection{Charged sleptons and sneutrinos}
\label{subsec:sleptons}

We focus first on the sneutrino sector as it shows two distinct
features compared to the MSSM. First, it gets enlarged by the
  superpartners of the right-handed neutrinos. Second,
 even
more drastically, splittings between the real and
 imaginary
 parts of
the sneutrinos occur resulting in 12 states: 6 scalar sneutrinos
and 6 pseudoscalar ones \cite{Hirsch:1997vz,Grossman:1997is}. The
origin of this splitting is the
$Y^{ij}_x\,\hat{\nu}_i\,\hat{\eta}\,\hat{\nu}_j$ in the
superpotential, \EQ~(\ref{eq:superpot}), which is a $\Delta L=2$
operator after the breaking of $U(1)_{B-L}$.
Therefore, we define
\begin{equation}
\tilde{\nu}^i_L = \frac{1}{\sqrt{2}}\left(\sigma^i_L + i \phi^i_L\right)\, ,
 \hspace{1cm} 
 \tilde{\nu}^i_R = \frac{1}{\sqrt{2}}\left(\sigma^i_R + i \phi^i_R\right)
 \,.
\end{equation}
The $6 \times 6$ mass matrices of the CP-even (\mReSnuSq) and
CP-odd (\mImSnuSq)
 sneutrinos can be written in the basis
\(\left(\sigma_{L},\sigma_{R}\right)\), respectively,
\(\left(\phi_{L},\phi_{R}\right)\) as
\begin{equation} 
\mReSnuSq = \Re\left( 
\begin{array}{cc}
m^R_{LL} &m^{R,T}_{RL}\\ 
m^R_{RL} &m^R_{RR}\end{array} 
\right), \hspace{1cm} 
\mImSnuSq = \Re\left( 
\begin{array}{cc}
m^I_{LL} &m^{I,T}_{RL}\\ 
m^I_{RL} &m^I_{RR}\end{array} 
\right) \, .\hspace{0.5cm} 
\end{equation} 
While $m^I_{LL}=m^R_{LL}=m_{LL}$ holds\footnote{We have neglected the
  splitting induced by the left-handed neutrinos as these are
  suppressed by powers of the light neutrino
 mass over the sneutrino mass.},
 the entries involving
``right-handed'' sneutrinos differ by a few signs.
 It is possible to
express them in a compact form by
\begin{align} 
 m_{LL} &=  m_{L}^{2} +\frac{v_u^2}{2} Y^\dagger_\nu  Y_{\nu} 
 + \frac{1}{8} \Big( 
    (g_{1}^{2}+g_{2}^{2}+\gmix ^{2} + \gmix  \gBL{} )
 (v_d^2- v_u^2)
   + 2 (\gBL{2} + \gmix  \gBL{} ) (v_{\eta}^2
 -  v_{\bar{\eta}}^2)\Big)
  {\bf 1} \, ,\\ 
 m^{R,I}_{RL} &= \frac{1}{\sqrt{2}} 
 \Big(v_u  T_{\nu}^{*} -  v_d \mu Y_{\nu}^{*} \Big)
      \pm  v_u v_{\eta} Y_x  Y_{\nu}^{*} \, ,\\ 
 \label{eq:mRR}
m^{R,I}_{RR} &= m_{{\nu}^{c}}^{2}
 +\frac{v_u^2}{2} Y_{\nu} Y_\nu^\dagger
+ 2 v^2_{\eta} Y_x Y_x^* \pm \sqrt{2} v_{\eta} T_x 
\mp \sqrt{2} Y_x  v_{\bar{\eta}} {\mu'}^*
 \nonumber \\ 
 & 
+ \frac{1}{8} \Big( 2 \gBL{2} (v^2_{\bar{\eta}}  - v^2_{\eta})
  + \gmix  \gBL{}  (v_{u}^{2} - v_{d}^{2})\Big) {\bf 1} \, .
\end{align} 
The upper signs correspond to the scalar and the lower ones to
the pseudoscalar matrices
 and we have assumed CP conservation.  In the case of
complex trilinear couplings or $\mu$ terms, a mixing between the
scalar and pseudoscalar particles occurs, resulting in 12 mixed states
and consequently in a $12\times 12$ mass matrix. In particular, the
term $\sim v_{\bar{\eta}} Y_x {\mu'}^*$ is potentially large and
induces a large mass splitting between the scalar and pseudoscalar
states. Also the corresponding soft SUSY-breaking term
 $\sim v_{\eta} T_x$ can
lead to a sizable mass splitting in the case of large $|A_0|$. 

The differences in the charged slepton sector compared to the MSSM
are additional D terms as well as a modification of the usual
D term. The mass matrix reads in the basis \( \left(\tilde{e}_{L},
 \tilde{e}_{R}\right) \) as
\begin{align} 
m^2_{\tilde{e}} &= \left( 
\begin{array}{cc}
m_{LL} & \frac{1}{\sqrt{2}} \Big(v_d T_{e}  - v_u \mu^* Y_{e} \Big)\\ 
\frac{1}{\sqrt{2}} \Big(v_d T^\dagger_{e}  - v_u \mu Y^\dagger_{e} \Big)
 & m_{RR}\end{array} 
\right) \, , \\ 
m_{LL} & =  m_{L}^{2} +\frac{v_d^2}{2} Y^\dagger_{e} Y_{e} + \frac{1}{8} \Big(
 (g_{1}^{2}  -g_{2}^{2} +\gmix ^{2}+\gmix  \gBL{} )
 (v_{d}^{2}- v_{u}^{2})
 +2(\gmix  \gBL{} + \gBL{2}) (v_{\eta}^{2}
  - v_{\bar{\eta}}^{2})\Big)
 {\bf 1} \, ,
\label{eq:mLL11}\\ 
m_{RR} & = m_{e^{c}}^{2}+\frac{v_d^2}{2} Y_{e} Y^\dagger_{e}
 + \frac{1}{8} \Big((2g_{1}^{2}+2\gmix ^{2}
+\gmix  \gBL{}  )
 (v_{u}^{2}- v_{d}^{2})-2\Big(2 \gmix  \gBL{} 
 + \gBL{2} \Big)(v_{\eta}^{2}
 - v_{\bar{\eta}}^{2}) \Big){\bf 1}  \, .
\end{align} 
For the first two generations one can neglect the left-right
mixing and  thus  the mass eigenstates
 correspond essentially to
the electroweak flavor eigenstates. In the following we will call the
 partners
of the left-handend (right-handed) leptons $L$ sleptons ($R$ sleptons).

\subsection{High-scale boundary conditions}

We will consider in the following a scenario motivated by minimal
supergravity, \EG we assume a GUT unification of all
soft SUSY-breaking scalar masses as well as a unification of all
 gaugino mass parameters
\begin{align}
 m^2_0 = & m^2_{H_d} = m^2_{H_u} = m^2_{\eta} = m^2_{\bar{\eta}} \,, \\
m^2_0 {\bf 1} =
 & m_{d^{c}}^{2} = m_{u^{c}}^{2} = m_Q^2 = m_{e^{c}}^{2} = m_{{\nu}^{c}}^{2}
 \,,\\
 M_{1/2} = & M_1 = M_2 = M_3 = M_{\tilde{B}'} \, .
\end{align}
Similarly, for the trilinear soft SUSY-breaking coupling the following
 \msugra  conditions are assumed:
\begin{align}
 T_i = A_0 Y_i, \hspace{1cm} i = e,d,u,x,\nu \thickspace . 
\end{align}

Furthermore, we assume that there are no off-diagonal gauge couplings
or gaugino mass parameters present at the GUT scale
\begin{align}
 g_{B Y} = & g_{Y B} = 0 \,,\\
 M_{B B'} = & 0 \, .
\end{align}
This choice is motivated by the possibility that the two Abelian
groups are a remnant of a larger product group that gets broken at
the GUT scale as stated in the Introduction.  In that case \(g_{YY}\)
and \(g_{B B}\) correspond to the physical couplings $g_1$ and
$\gBL{}$, which we assume to unify with $g_2$:
\begin{equation}
 g^{GUT}_1 = g^{GUT}_2 = \gBL{GUT} \thickspace ,
\end{equation}
where we have taken into account the GUT normalization
discussed in Sec. \ref{subsec:kineticmixing}.

In addition, we take the mass of the $Z'$ and $\tan\beta'$ as
inputs and use the following set of free parameters
\begin{eqnarray}
& m_0, \thickspace M_{1/2},\thickspace A_0,\thickspace \tan\beta,\thickspace
 \tan\beta',\thickspace \sign(\mu),\thickspace \sign(\mu'),\thickspace M_{Z'},
 \thickspace  Y_x \thickspace \mbox{and} \thickspace Y_{\nu}\, . &
\end{eqnarray}
\(Y_{\nu}\) is constrained by neutrino data and must therefore be very
small compared to the other couplings,
 \EG they are of the order of the electron Yukawa coupling.
Therefore,  they  can be safely neglected
in the following. $Y_x$ can always be taken diagonal and thus
effectively we have nine free parameters and two signs.

\section{Numerical results}
\label{sec:results}

\subsection{Implementation in \SARAH and \SPheno}

All analytic expressions for masses, vertices, RGEs, as well as
one-loop corrections to the masses and tadpoles were calculated using
the \SARAH package \cite{Staub:2008uz,Staub:2009bi,Staub:2010jh}.  
The RGEs are included at the two-loop level 
in the most general form respecting the complete flavor structure
using the formulas of \REF{Martin:1993zk} augmented
by gauge kinetic mixing effects as discussed in 
 \REF{Fonseca:2011vn}. The RGEs and
 the loop corrections to
all masses as well as to the tadpoles are derived in
 the \DRbar\ scheme
and the Feynman-'t Hooft gauge.

The numerical evaluation of the model is very similar to that of the
default implementation of the MSSM in \SPheno
\cite{Porod:2003um,Porod:2011nf}: as the starting point, the SM gauge
and Yukawa couplings are determined using one-loop relations, as given
in ref.~\cite{Pierce:1996zz}, that are extended to our model.  The
vacuum expectation values \(v_d\) and \(v_u\) are calculated with
respect to the given value of \(\tan\beta\) at \(M_Z\), while
\(v_{\eta}\) and \(v_{\bar{\eta}}\) are derived from the input values
of $M_{Z'}$ and $\tan\beta'$ at the SUSY scale.

The RGEs for the gauge and Yukawa couplings are evaluated up to the
SUSY scale, where the input values of $Y_{\nu}$ and $Y_x$ are set.
Afterwards, a further evaluation of the RGEs up to the GUT scale takes
place. 
After setting the boundary conditions, all parameters are evaluated
back to the SUSY scale. There, the one-loop-corrected SUSY masses are
calculated using on-shell external momenta. These steps are iterated
until the relative change of all masses between two iterations is
below $10^{-4}$. 

We have used {\tt WHIZARD} \cite{Kilian:2007gr} to evaluate 
the bounds on the $Z'$ discussed in Sec.~\ref{subsec:pheno} 
as well as for the signals for slepton production
in Sec.~\ref{subsec:sleptonprod}. For this purpose, the
{\tt SUSY Toolbox} \cite{Staub:2011dp} has been used to implement 
 the  model in  {\tt WHIZARD}
  based on the corresponding model files written by {\tt SARAH} and to
perform the parameter scans with {\tt SSP}.

\subsection{Parameter studies}

\begin{table}
\center

\begin{tabular}{| l | c  c|} \hline
  & ~~~BLV~~~ & ~~~BLVI~~~ \\ \hline\hline
  \multicolumn{3}{|c|}{Input parameters} \\ \hline
$m_0 ~[\text{TeV}]$ & 1 & 0.6 \\
$M_{1/2} ~[\text{TeV}]$ & 1.5 & 0.6 \\
$A_0~[\text{TeV}]$ & -1.5 & 0 \\
$\tan \beta$ & 20 & 10 \\
sign$~\mu$ & + & + \\
$\tan\beta'$ & 1.15 & 1.07 \\
sign$~\mu'$ & + & + \\
$\mzp~[\text{TeV}]$ & 2.5 & 2 \\
$Y_X^{11}$ & 0.37 & 0.42 \\
$Y_X^{22}$ & 0.4 & 0.43 \\
$Y_X^{33}$ & 0.4 & 0.44 \\ \hline
\end{tabular}
\hspace{0.5cm}
\begin{tabular}{| l | c  c |}\hline
  & ~~~BLV~~~ & ~~~BLVI~~~ \\ \hline\hline
\multicolumn{3}{|c|}{Masses [GeV]} \\ \hline
$m_{\tilde \chi^0_1}~~~~$  & 678.0 & 280.7\\
$m_{\tilde \chi^0_2}$  & 735.2 & 475.4\\
 \hline
$m_{\tilde \chi^\pm_1}$  & 1242.0 & 475.4\\
\hline
$m_{\tilde \tau_1}$  & 1002.0 & 603.7\\
$m_{\tilde \tau_2}$  & 1446.5 & 759.9\\
$m_{\tilde \mu_{R}}$  & 1094.2 & 610.8\\
$m_{\tilde \mu_{L}}$  & 1477.4 & 761.9\\
$m_{\tilde e_{R}}$  & 1094.5 & 610.8\\
$m_{\tilde e_{L}}$  & 1477.5 & 761.9\\ \hline
$m_{\tilde \nu_1^R}$  & 811.3 & 754.9\\
$m_{\tilde \nu_1^I}$  & 1442.4 & 754.9\\ \hline
\end{tabular}
\caption{Parameters of the study points and selected masses.
}
\label{tab:parameter_points}
\end{table}

\begin{figure}
\center
\begin{minipage}[b]{8.1cm}
\includegraphics[height=5.4cm]{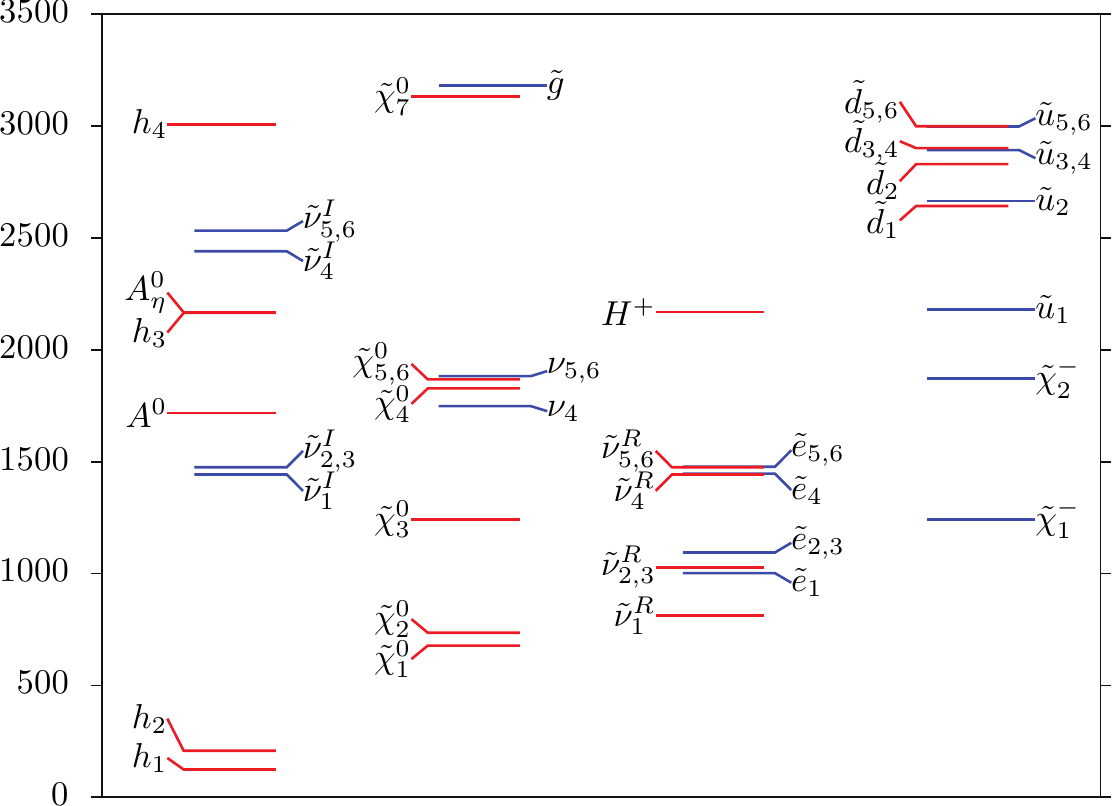}
\end{minipage}
\begin{minipage}[b]{8.1cm}
\includegraphics[height=5.4cm]{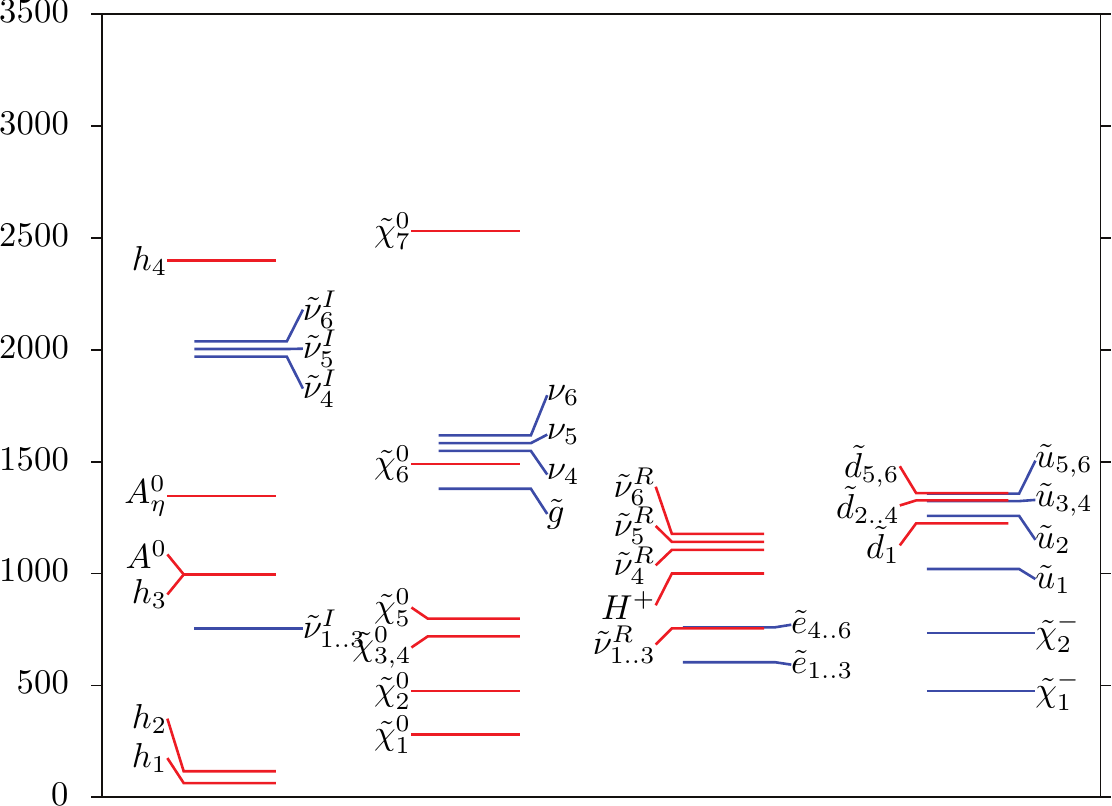}
\end{minipage}
\caption{The mass spectra of BLV (left) and BLVI (right).}
\label{fig:BLV_and_BLVI_spectra}
\end{figure}

\begin{figure}
\center
\begin{minipage}[b]{8.1cm}
\includegraphics[height=7.4cm]{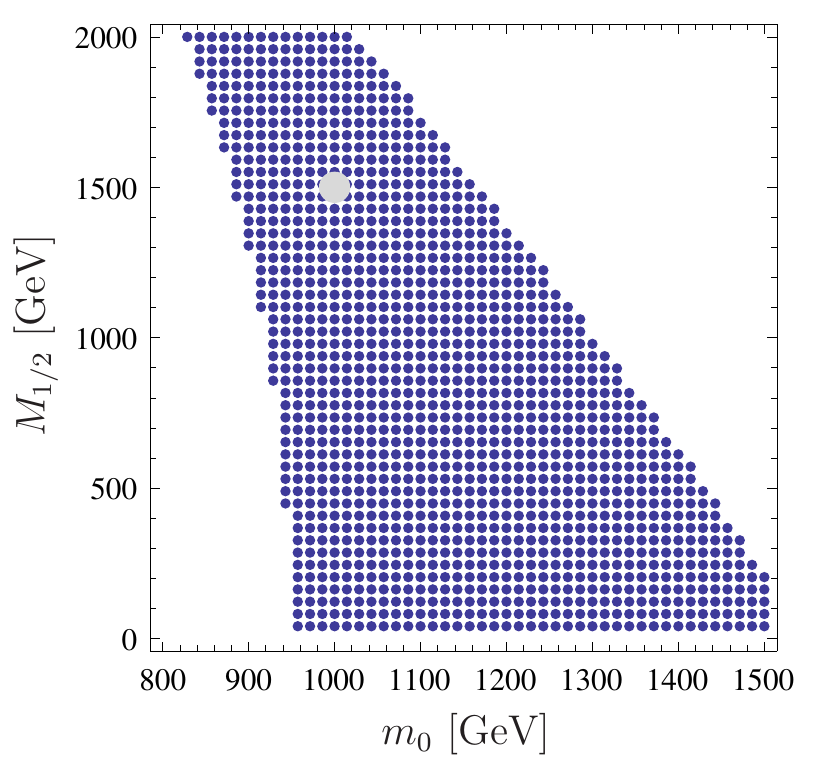}
\end{minipage}
\begin{minipage}[b]{8.1cm}
\includegraphics[height=7.4cm]{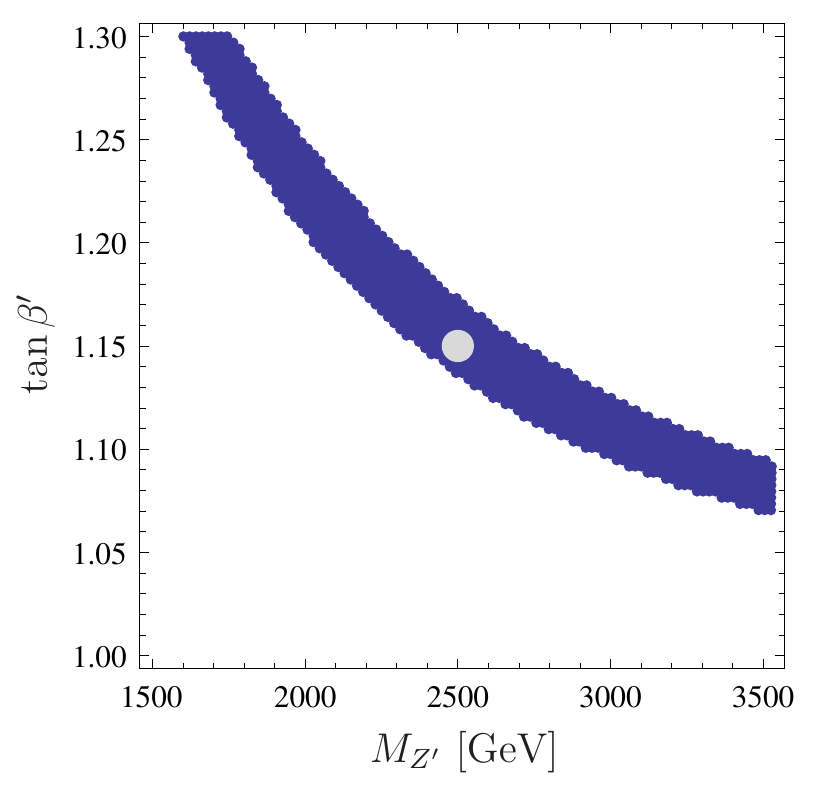}
\end{minipage}
\caption{ Allowed regions in the $m_0-M_{1/2}$ plane for 
$\mzp=2.5~\text{TeV},~\tan\beta'=1.15$ (left) and in 
the $\mzp-\tan\beta'$ 
plane for $m_0=1$~TeV and $M_{1/2}=1.5$~TeV (right).
In both plots we have fixed in addition  
$\tan \beta=20$ and $A_0=-1.5$~TeV.
The grey dot indicates the selected benchmark point (BLV).}
\label{fig:MZpTBpScan3}
\end{figure}

\begin{figure}
\center
\begin{minipage}[b]{8.1cm}
\includegraphics[height=7.4cm]{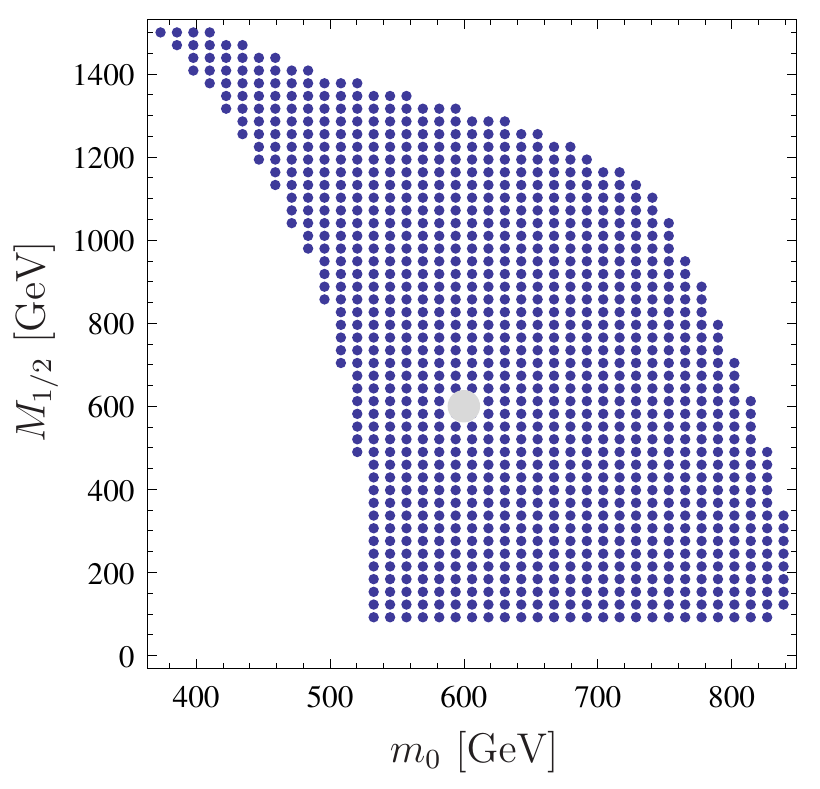}
\end{minipage}
\begin{minipage}[b]{8.1cm}
\includegraphics[height=7.4cm]{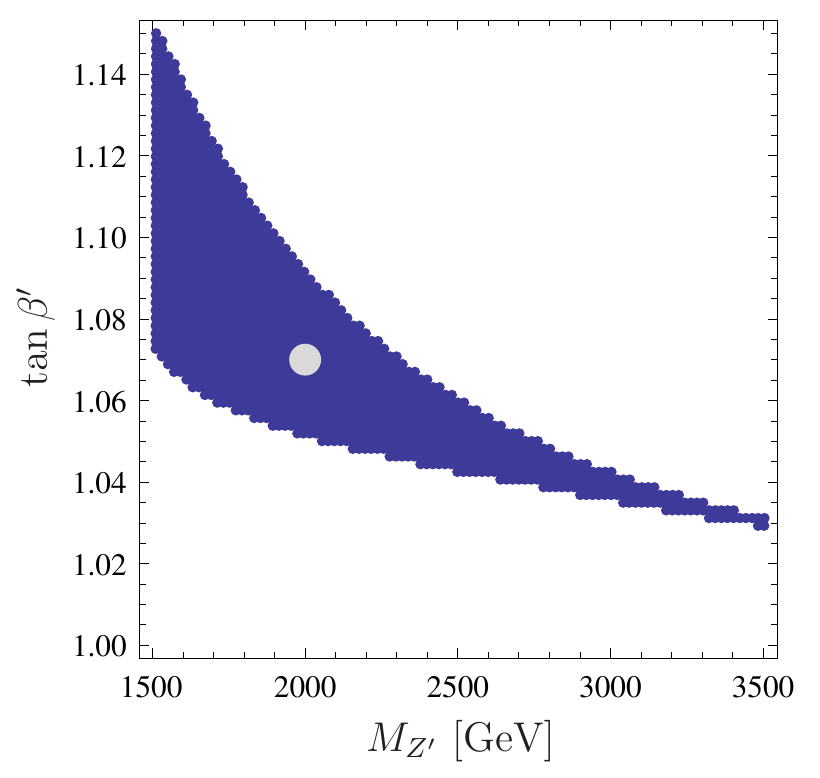}
\end{minipage}
\caption{Allowed regions in the $m_0-M_{1/2}$ plane for 
$\mzp=2~\text{TeV},~\tan\beta'=1.07$ (left) and in 
the $\mzp-\tan\beta'$ 
plane for $m_0=0.6$~TeV and $M_{1/2}=0.6$~TeV (right).
In both plots we have fixed in addition  
$\tan \beta=10$ and $A_0=0$.
The grey dot indicates the selected benchmark point (BLVI).}
\label{fig:MZpTBpScan}
\end{figure}

In all numerical evaluations we have used the following SM input:
$G_F = 1.6639 \cdot 10^{-5} $~GeV$^{-2}$, $m_Z=91.187$~GeV,
 $m_\tau=1.7771$~GeV,
$m_t = 172.9$~GeV, $m_b (m_b) = 4.2$~GeV and $\alpha_s(m_Z)= 0.119$.
The latter two are 
$\overline{MS}$ values that are converted to the $\overline{DR}$
scheme.  Moreover, we have fixed the neutrino Yukawa couplings
 such that the light neutrino masses can be explained. As
this requires the maximum of $|Y_{\nu,ij}|$ to be at most
$10^{-5}$, these couplings do not play any role in the considerations
here. Because of their smallness, one can automatically satisfy the
constraints from the nonobservation of rare decays such as 
$\mu\to e\gamma$ over all the parameter space. This is 
in contrast to the usual MSSM augmented
by the various seesaw mechanisms, as discussed in \EG 
\cite{Esteves:2009vg,Esteves:2010ff} and
references therein.

As a starting point for our numerical investigations,
we have taken the two points defined in \TAB~\ref{tab:parameter_points}.
In analogy to \REF{O'Leary:2011yq}, we label them BLV and BLVI (see \FIG \ref{fig:BLV_and_BLVI_spectra}).
BLVI has a spectrum close to the existing LHC exclusion
bounds. Note, however, that the high-scale input corresponds
to different SUSY particle masses in comparison to the CMSSM
because the running of the parameters is not the same 
and also the mass matrices differ. In contrast, BLV
 has  quite a heavy spectrum and can only be discovered at
 $\sqrt{s}=14$~TeV. As discussed above, in this model 
the extra gauge group implies additional
D terms to the sfermion masses. The requirement
that the mass-squared parameters for all
 the sfermions are positive restricts
the allowed range for $\tan\beta'$, as can be seen
in \FIGS~\ref{fig:MZpTBpScan3} and \ref{fig:MZpTBpScan}.
This is  a consequence of the large mass for the $Z'$. Clearly this
 restriction is more severe for large $M_{Z'}$
and less severe for 
large values of $m_0$ and $M_{1/2}$.

For completeness, we note that these points would give too
large a relic density, which however is not an insurmountable problem and
can easily be fixed without changing the collider phenomenology.
 In case of BLV one would need
to invoke nonuniversal boundary conditions for the bileptons
to change its mass. In this way one
obtains an efficient annihilation of the lightest neutralino
via a bilepton resonance as discussed in \cite{Basso:2012gz}
but having at the same time only tiny effects
on the various decay branching ratios.
In case of the BLVI the addition of a  nonthermally produced
gravitino with a mass of 
10 GeV gives the correct relic density yielding a lifetime for
the neutralino of about $10^{-3}$ seconds. This is sufficiently
long-lived to appear as a stable particle at the LHC and at the
same time sufficiently short-lived so that there are no problems
with big bang nucleosynthesis.

\subsection{$Z'$ phenomenology}
\label{subsec:pheno}

The mass of additional vector bosons as well as their mixing with
the SM $Z$ boson, which implies, for example, a deviation of the fermion
couplings to the $Z$ boson compared to SM expectations, is severely
constrained by precision measurements from the LEP experiments
\cite{Alcaraz:2006mx,Erler:2009jh,Nakamura:2010zzi}.
The bounds are on both the mass of the $Z'$ and the mixing with
the standard model $Z$ boson, where the latter is constrained by
$|\sin(\Theta_{W'})<0.0002|$. Using \EQ~(\ref{eq:ThetaWP}) together with
\EQ~(\ref{eq:MassZP}) as well as the values of the running gauge couplings,
a limit on the $Z'$ mass of about 1.2~TeV is obtained. Taking in addition 
the bounds obtained from $U$, $T$ and $S$ parameters into account
\cite{Cacciapaglia:2006pk} one gets $M_{Z'} / ( Q_e^{B-L} \gBL{} ) > 6.7$~TeV
 that for $\gBL{} \simeq 0.55$ would imply $M_{Z'} \gsim 1.84$~TeV. 
However, this coupling has to be replaced by the effective
coupling that gets modified due to gauge kinetic mixing:
see Appendix~\ref{sec:couplings}\footnote{Strictly
speaking, this effect modifies the couplings to left- and right-handed
fermions differently. Taking this into account would require a 
re-evaluation of the complete analysis which is beyond the scope
of this paper. To be conservative we have taken the larger of
the two couplings.}. Therefore, the above formula reads in
our model $M_{Z'} / ( Q_e^{B-L} ( \gBL{} + \gmix ) ) > 6.7$~TeV,
and $\gBL{} \simeq 0.55, \gmix \simeq - 0.11$
imply $M_{Z'} \gsim 1.47$~TeV.

\begin{figure}[t]
\centering
\begin{minipage}[b]{12cm}
\includegraphics[width=12cm]{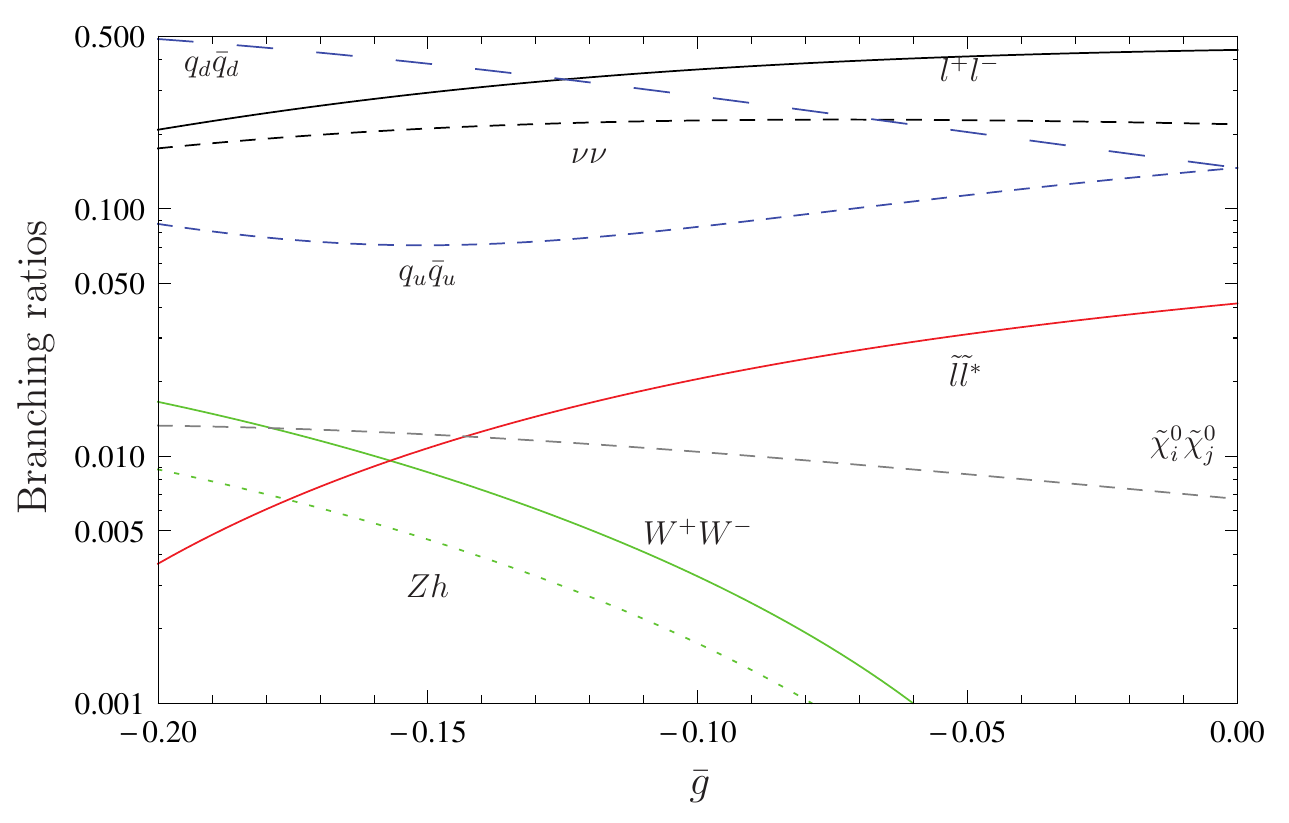}
\end{minipage}
\begin{minipage}[b]{12cm}
\includegraphics[width=12cm]{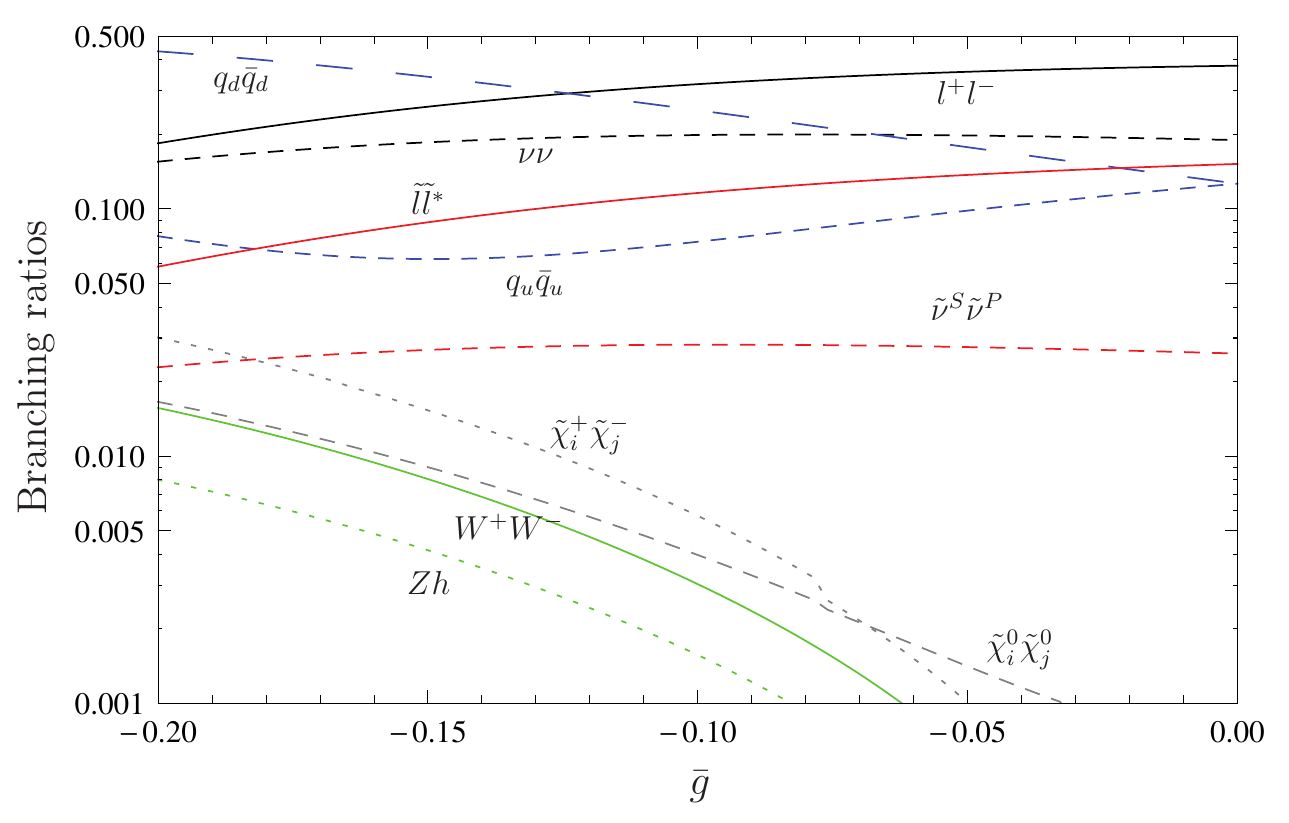}
\end{minipage}
\caption{Branching ratios of the $Z'$ at the considered parameter points
 BLV
 (above) and BLVI (below) as a function of the off-diagonal coupling
 parameter \gmix.}
\label{fig:BRs_versus_gYB}
\end{figure}

\begin{figure}[t]
\begin{minipage}[b]{8.1cm}
\includegraphics[height=5cm]{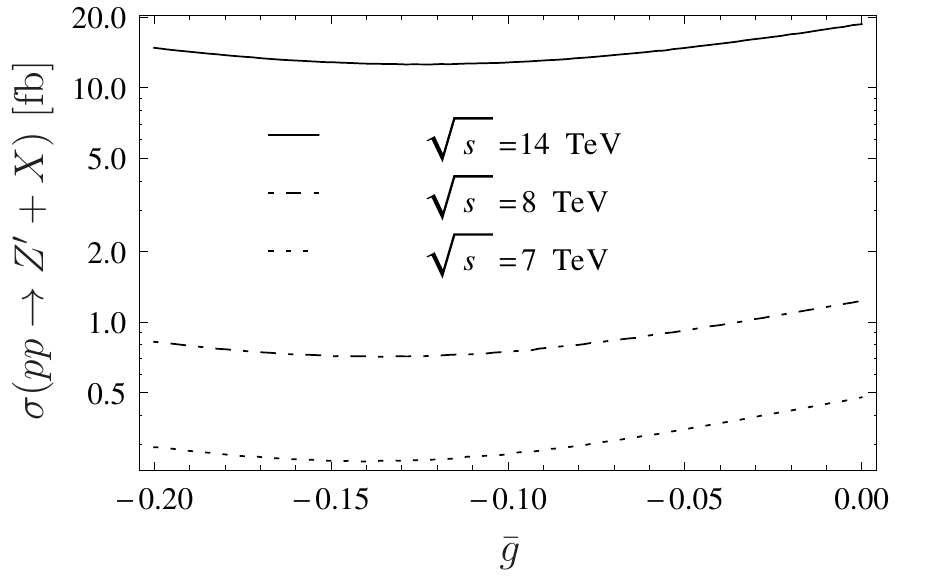}
\end{minipage}
\begin{minipage}[b]{8.1cm}
\includegraphics[height=4.9cm]{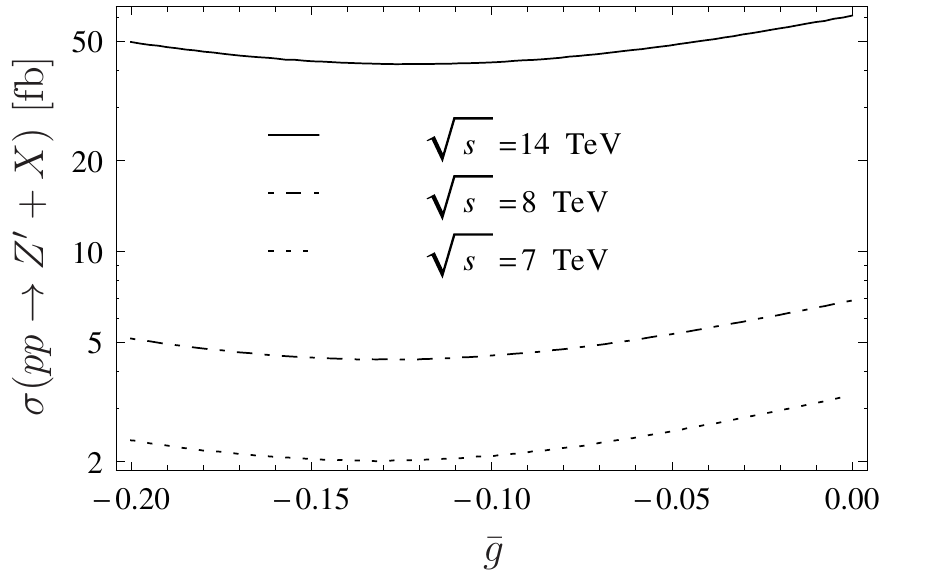}
\end{minipage}
\caption{LHC production cross sections of the $Z'$ at the considered 
parameter points BLV (left) and BLVI (right) and three different
 center-of-mass 
energies as a function of the off-diagonal coupling parameter \gmix.}
\label{fig:sigma_versus_gYB}
\end{figure}

The $Z'$ dominantly decays into SM fermions as can be seen
in \FIG~\ref{fig:BRs_versus_gYB} where we show the
branching ratios as a function of \gmix. We have fixed
all other parameters as given for the two study points in
\TAB~\ref{tab:parameter_points}.
For the study points themselves we find $\gmix \simeq -0.11$,
which has to be compared with $\gBL{} \simeq 0.55$. As they are
of the same order of magnitude, one can easily understand
the strong dependence of the various branching ratios after
inspecting the couplings given in Appendix~\ref{sec:fcouplings}.
The $Z'$ can also decay into supersymmetric particles 
\cite{Gherghetta:1996yr} with branching ratios of up to $O(10\%)$
in our model. We find that, in particular, decays into charged sleptons
can have a sizable branching ratio, as can also be seen in
\FIG~\ref{fig:BRs_versus_gYB}. Besides the decays,
the cross sections also depend on gauge kinetic mixing as
demonstrated in \FIG~\ref{fig:sigma_versus_gYB} where
we show the $Z'$ cross section for $\sqrt{s} = 7,8$, and 14~TeV
as a function of \gmix. For the PDFs we have used the set
 \texttt{CTEQ6L1} \cite{cteq}.

\begin{table}[t]
\center
\begin{tabular}{|c| c c|}\hline
 Parameter point & $\gmix = -0.11$ & $\gmix = 0$ \\ \hline \hline
BLV & 1770~GeV & 1965~GeV \\
BLVI & 1730~GeV & 1900~GeV  \\ \hline
\end{tabular}
\caption{Current bounds on $\mzp$ in the supersymmetric $B-L$ model
 derived from 5~fb${}^{-1}$ of ATLAS data for both benchmark points using
 different assumptions as discussed in the text.}
\label{tab:bounds_table}
\end{table}
\begin{figure}[t]
\centering
\begin{minipage}[b]{8.1cm}
\includegraphics[width=8.1cm]{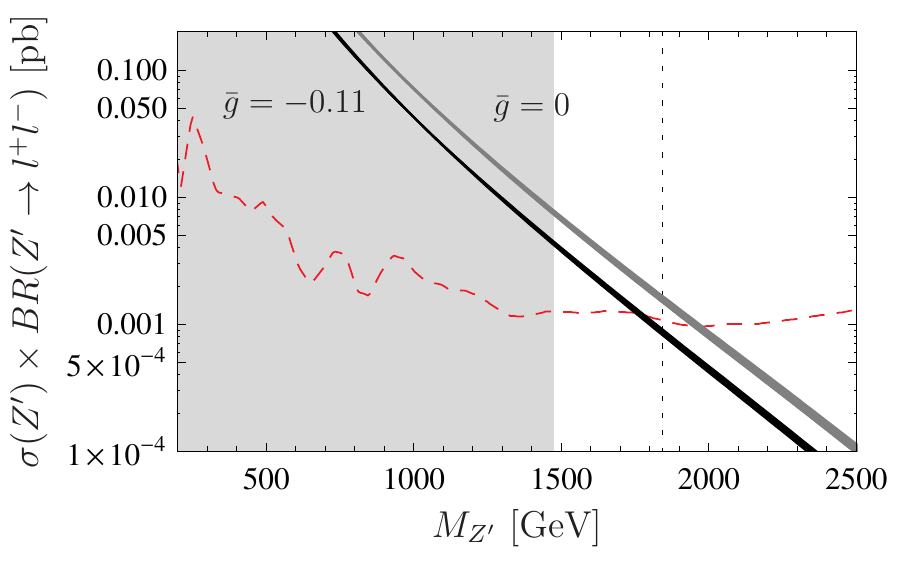}
\end{minipage}
\begin{minipage}[b]{8.1cm}
\includegraphics[width=8.1cm]{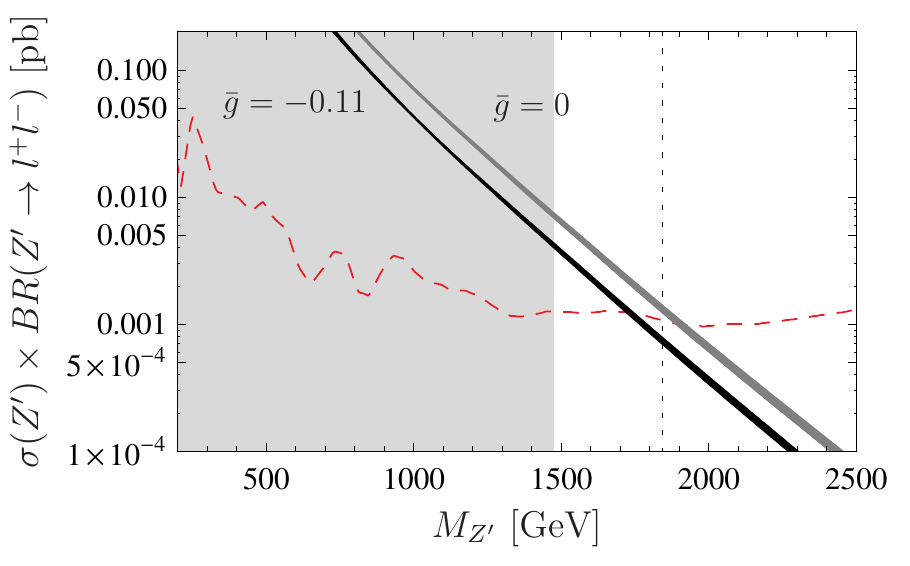}
\end{minipage}
\caption{Current limits on $\mzp$ for combined lepton production
for the study points BLV (left) and BLVI (right). The red dashed curve
 shows the 
recent experimental ATLAS limits  \cite{Collaboration:2011dca}. 
The black and grey bands are the dilepton production
 cross sections at the 
$Z'$ peak for the case of $U(1)$ mixing (black) and without
 (grey). The 
grey shaded area shows the mass range forbidden by LEP II, while the black 
dotted line shows the LEP limits without taking into account gauge kinetic 
mixing.}
\label{fig:lhc_limits}
\end{figure}

Recently ATLAS and CMS \cite{Collaboration:2011dca,Chatrchyan:2011wq}
have updated the results for $Z'$ searches. As they do not give
direct bounds for our model, we have calculated the corresponding
signal cross section.
If one took only final states containing SM particles
into account and neglected gauge kinetic mixing, one would find
a lower bound of $ 1970~\text{GeV} \le M_{Z'}$. Taking into account
gauge kinetic mixing we find for $\gmix \simeq -0.1$ a bound
of 1790~GeV.
Moreover, for both our study points, final states 
containing supersymmetric particles are also present as discussed above. 
This leads to an increase of the
width and thus to a reduction of the signal
 cross sections.
Therefore the bounds are less severe, as has also been
 discussed 
in the context of related models 
\cite{Chang:2011be,Corcella:2012dw}. In contrast
to previous studies, we encounter here a case where neglecting
 gauge kinetic effects would lead to a significantly incorrect bound.
In fact the bound obtained at LHC differs by about 200~GeV depending 
on whether 
kinetic mixing is correctly taken into account or not,
as can be seen from \TAB~\ref{tab:bounds_table}. This
is further exemplified in \FIG~\ref{fig:lhc_limits}  
where we display the cross section into leptons
(summed over electrons and muons)
as a function of $M_{Z'}$. The grey area is excluded 
by precision LEP data whereas the red dashed line gives the bound on
the signal cross section as obtained by the ATLAS Collaboration
\cite{Collaboration:2011dca}. Note that the effect
on the LEP limits is even stronger, as can be seen at the dotted line
 in \FIG~\ref{fig:lhc_limits}. 
The CMS bounds are similar and thus lead to nearly the same limits. 
Clearly it would be desirable to have a 
combined  analysis by both collaborations.

\begin{figure}[t]
\centering
\begin{minipage}[b]{8.1cm}
\includegraphics[height=5cm]{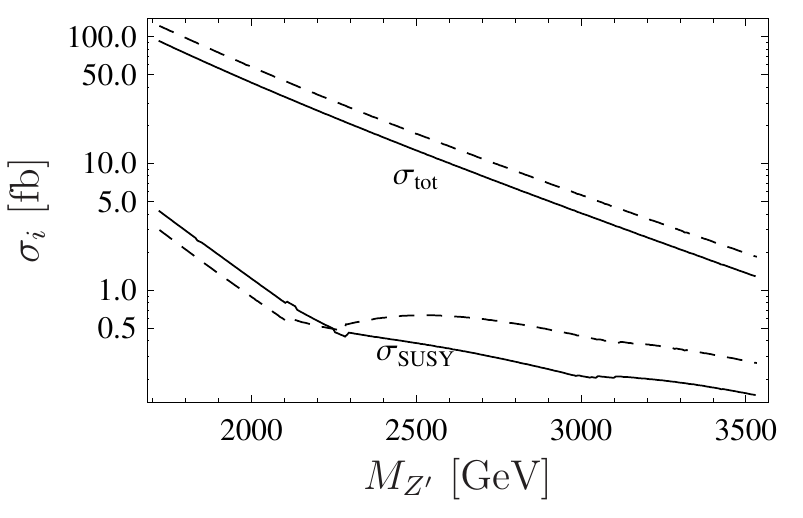}
\end{minipage}
\begin{minipage}[b]{8.1cm}
\includegraphics[height=5cm]{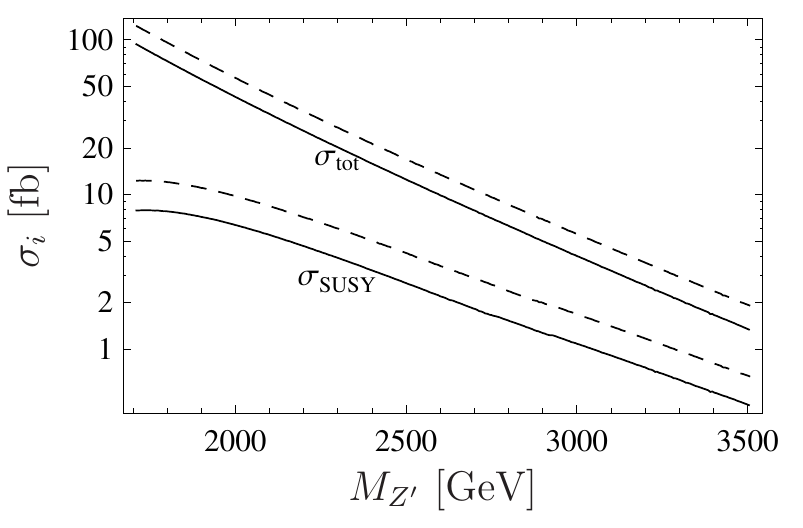}
\end{minipage}
\caption{Cross sections for $pp \to \zp$ at LHC with
$\sqrt{s}=$14~TeV as a function of $\mzp$ for the benchmark points BLV
 (left)
and BLVI (right). The upper two lines show the total cross section
 whereas
the lower two show the cross section in SUSY particles via a $Z'$.
The solid lines give the cross section taking into account gauge 
kinetic mixing whereas the dashed ones give the cross section
if gauge kinetic mixing were neglected.}
\label{fig:cross_section}
\end{figure}

In \FIG~\ref{fig:cross_section} we give the cross sections for $pp \to \zp$ 
at LHC with
$\sqrt{s}=$14~TeV as a function of $\mzp$ for the two benchmark points. 
To demonstrate the importance of gauge kinetic mixing, we display
the values taking it into account (solid lines) and
those where it is neglected (dashed lines). Comparing
both data points, one sees that the dependence on the gauge kinetic
mixing on the SUSY final states depends on the underlying soft
 SUSY-breaking parameters.  For BLVI, the dominant SUSY channels always
 involve the sleptons and therefore kinetic mixing reduces
 the cross section. 
In contrast, for BLV and small values of $M_{Z'}$, final 
states including two neutralinos with large bileptino contents
are important. Therefore, the cross section is larger with kinetic mixing
 than without. However, the masses of these neutralinos
rapidly increase with $M_{Z'}$ and slepton production is the dominant
SUSY channel for $M_{Z'} > 2.3$~TeV.

\subsection{Slepton production via $Z'$ as discovery channel}
\label{subsec:sleptonprod}

\begin{figure}[t]
\centering
\includegraphics[width=10cm]{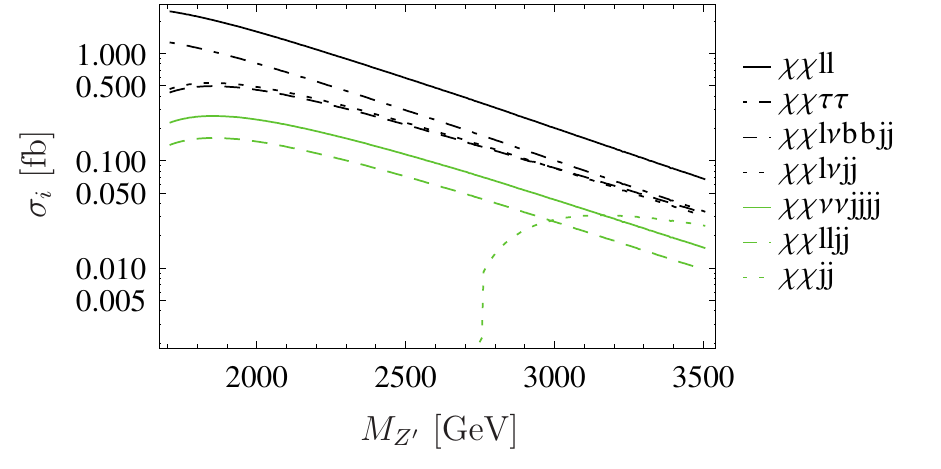}
\caption{Cross sections for the dominant final states resulting from
the SUSY particles that are produced via a $Z'$ for benchmark point
 BLVI.
The symbols correspond
to $\chi \mathrel{\widehat{=}} \tilde \chi^0_1$, l $\mathrel{\widehat{=}}$
(anti)lepton ($e,\mu$), b $\mathrel{\widehat{=}}$ (anti)bottom quark,  
$\nu \mathrel{\widehat{=}}$ neutrino, j  $\mathrel{\widehat{=}}$ jet
resulting from a quark of the first two generations.}
\label{fig:biggest_channels}
\end{figure}

\begin{figure}[t]
\centering
\includegraphics[height=4cm]{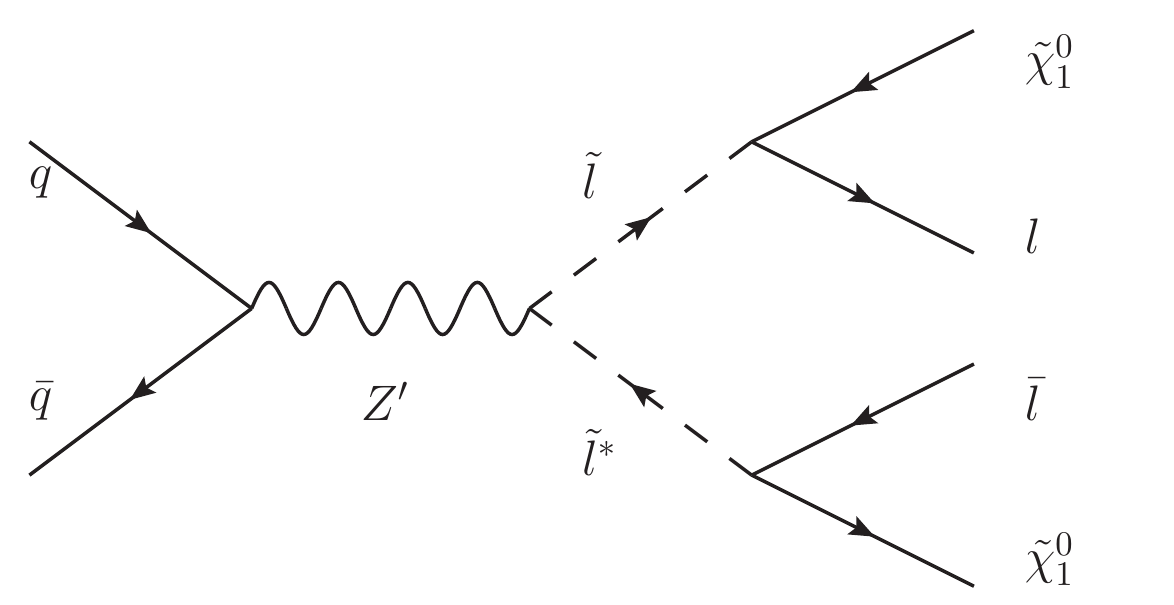}
\caption{Feynman diagram for the production of
$\tilde l \tilde l^*$ via an $s$-channel \zp   exchange and the subsequent
 decay into $l^+l^-\tilde \chi^0_1 \tilde \chi^0_1$.}
\label{feynman_mumuNN}
\end{figure}

A heavy $Z'$ allows the production of electroweak
SUSY particles with masses of several hundred GeV implying
that this might be an important discovery channel
 as has also been discussed in related models 
\cite{Kang:2004bz,Chang:2011be,Corcella:2012dw,Baumgart:2006pa}. 
In contrast to the previous studies, gauge kinetic mixing 
is important as we have seen above. In the following
we will first take benchmark point BLVI to discuss some basic
features. Much as in the MSSM, the $R$ sleptons
 decay almost always into a lepton and the
 lightest neutralino, whereas the $L$ sleptons
decay dominantly into the lighter chargino and a neutrino, or 
the second lightest neutralino and a
 lepton, in roughly the ratio two to one. The
 lighter chargino and the second lightest neutralino, which have a
 large wino fraction, decay dominantly into the
lightest neutralino and a vector boson.
The neutralino decay can also result in a final-state
 Higgs boson. Therefore one has additional
leptons and jets stemming from the 
decays of the vector boson and Higgs bosons. 

In \FIG~\ref{fig:biggest_channels} we display the most important
final states resulting from the cascade decays of supersymmetric
particles originating from the $Z'$. We have fixed the soft
 SUSY-breaking parameters as in BLVI. Nevertheless,
 the masses change as a function
of $M_{Z'}$ as the corresponding bilepton \vevs enter 
the mass matrices,
see \EG Secs. \ref{subsec:neutralinos} and \ref{subsec:sleptons}.
The dominant final state contains two leptons and two lightest
 supersymmetric particles stemming from slepton production as 
 displayed in \FIG~\ref{feynman_mumuNN}.
 For these points, sleptons are hardly ever produced in the
 cascade decays
of squarks and gluinos, in contrast to the neutralinos and charginos.

\begin{figure}[t]
\centering
\begin{minipage}[b]{13cm}
\includegraphics[width=13cm]{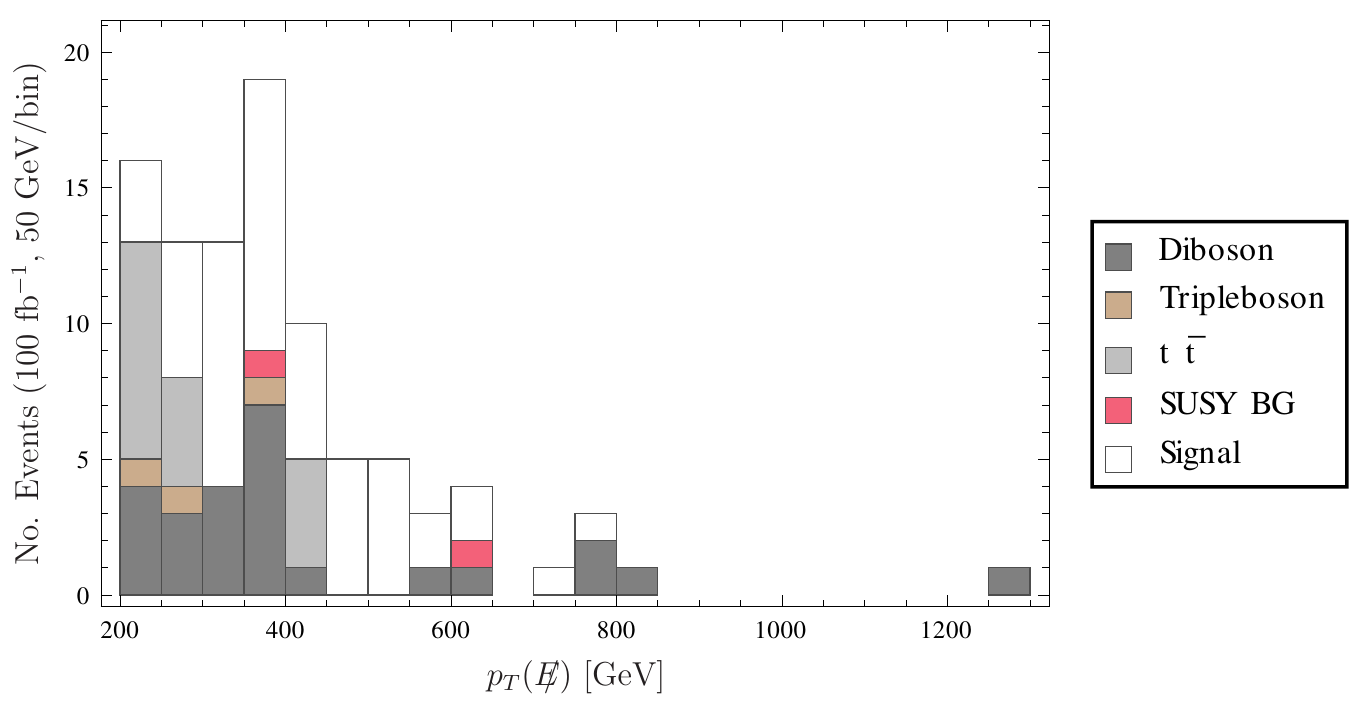}
\end{minipage}
\begin{minipage}[b]{13cm}
\includegraphics[width=13cm]{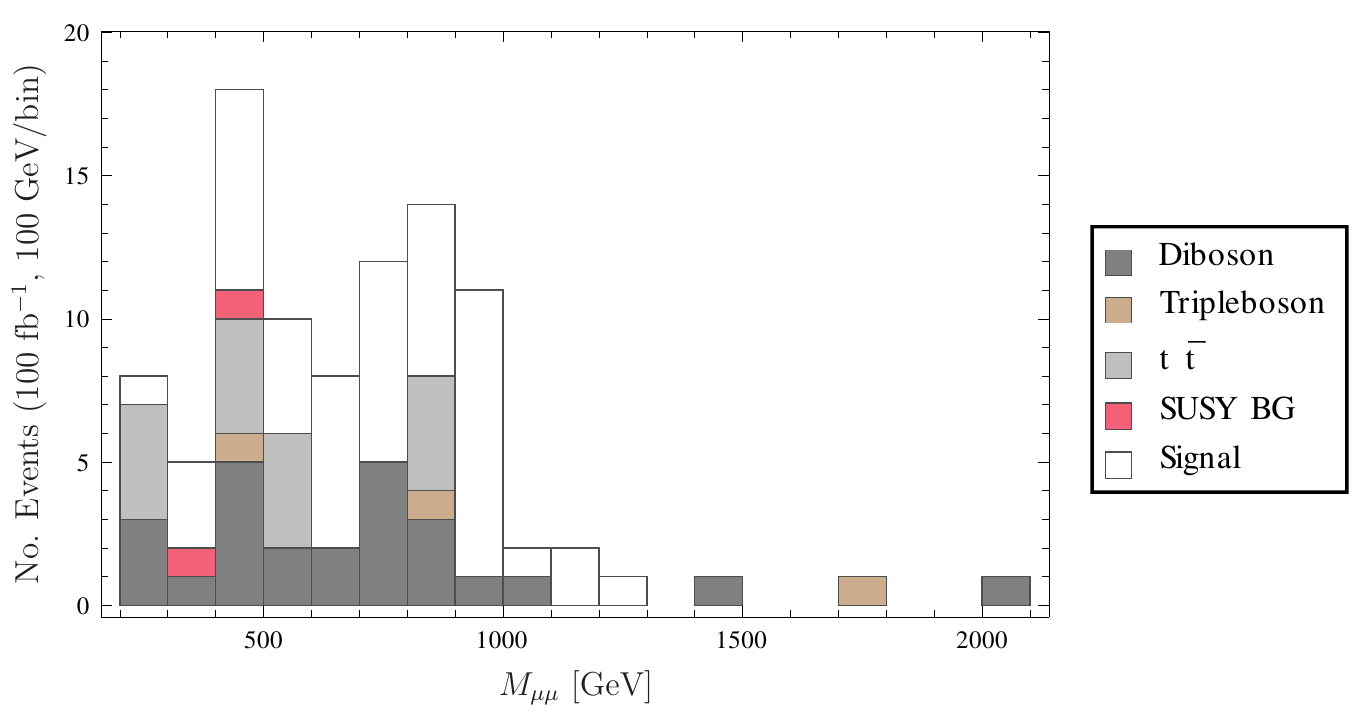}
\end{minipage}
\begin{minipage}[b]{13cm}
\includegraphics[width=13cm]{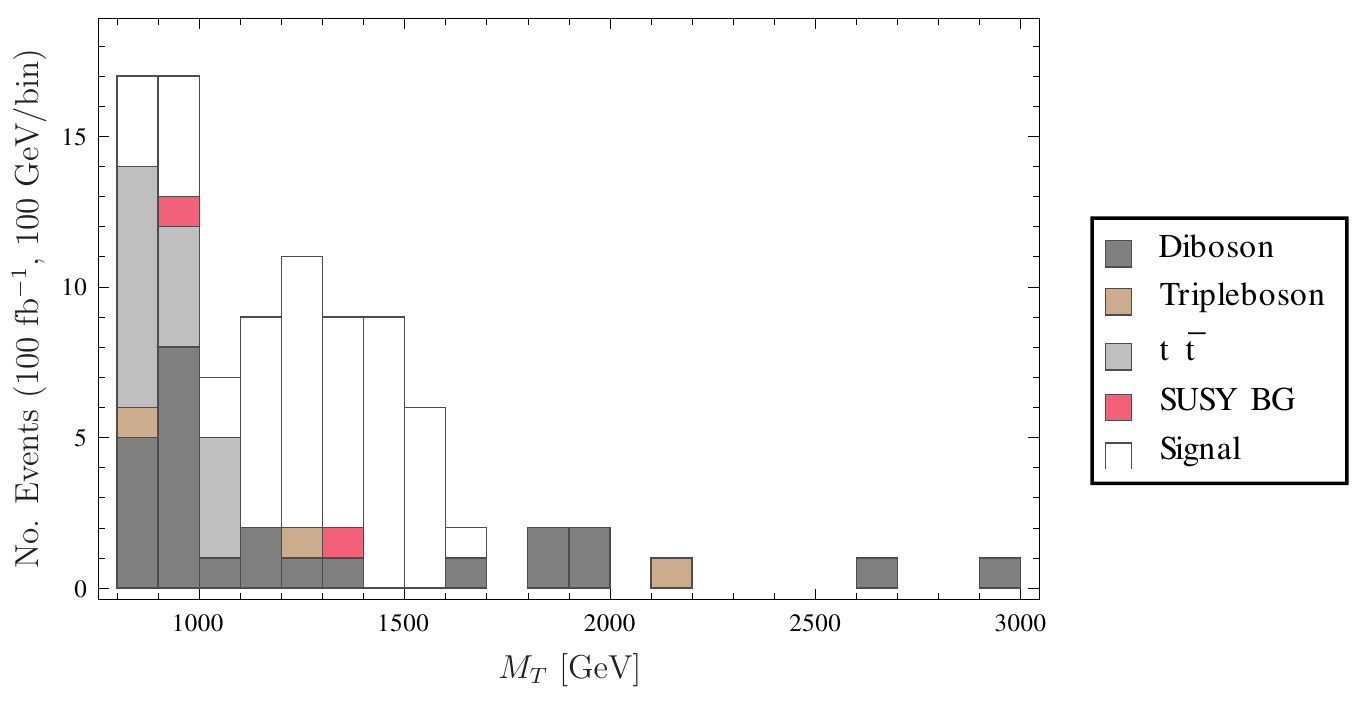}
\end{minipage}
\caption{Histograms of the $\mu^+ \mu^-+\emiss$ production with the applied 
cuts $M_{\mu \mu} > 200$~GeV, $p_T(\emiss) > 200$~GeV and $M_T > 800$~GeV 
as a function of the missing energy, muon pair
 invariant  mass and the transverse cluster mass.}
\label{fig:histos_withCuts}
\end{figure}

For this reason the $Z'$ decays are a potential discovery channel
for sleptons. We have performed a basic Monte Carlo study using
{\tt WHIZARD} \cite{Kilian:2007gr} to generate the signal in the
case of smuon pair production and simulated
the background. For the background, we considered
 diboson production, 
triple vector boson production and $t\bar t$ production as well
as neutralino and chargino production including both 
direct production via Drell-Yan processes and cascade decays from
squarks and gluinos. However, the contributions from the latter
are rather small.
We have applied the following cuts to suppress the background
\begin{itemize}
\item Invariant mass of the muon pair:  $M_{\mu \mu} > 200$~GeV.
\item Missing transverse momentum: $p_T(\emiss) > 200$~GeV.
\item A cut on the transverse cluster mass
\begin{align}
M_T = \sqrt{\left( \sqrt{p_T^2(\mu^+\mu^-) + M_{\mu\mu}^2}
 + p_T(\emiss) \right)^2
 - \left( \vec p_T(\mu^+\mu^-)+\vec p_T (\emiss) \right)^2}
\label{eq:m_T_1}
\end{align}
 with $\vec p_T$ being the 2D vector of the transverse momentum and
 $\vec p_T(\mu^+ \mu^-) = \vec p_T(\mu^+)+\vec p_T(\mu^-)$. We required
 $M_T > 800$~GeV.
\item For the suppression of $t\bar{t}$ and squark/gluino
 cascade decays,
we set a cut on the transverse momentum of the hardest jet:
$p_{T,jet} < 40$~GeV. 
\end{itemize}
In \FIG~\ref{fig:histos_withCuts} we display the resulting
distributions for an integrated luminosity of 100 fb$^{-1}$ at
$\sqrt{s}=14$~TeV. The resulting significance, which is calculated as
\begin{equation}
s = \frac{N_\text{Signal}}{\sqrt{N_\text{BG}}}\, ,
\end{equation}
is 7.5~$\sigma$, which is sufficient to claim discovery for
 this  example.
\begin{figure}[t]
\centering
\includegraphics[width=10cm]{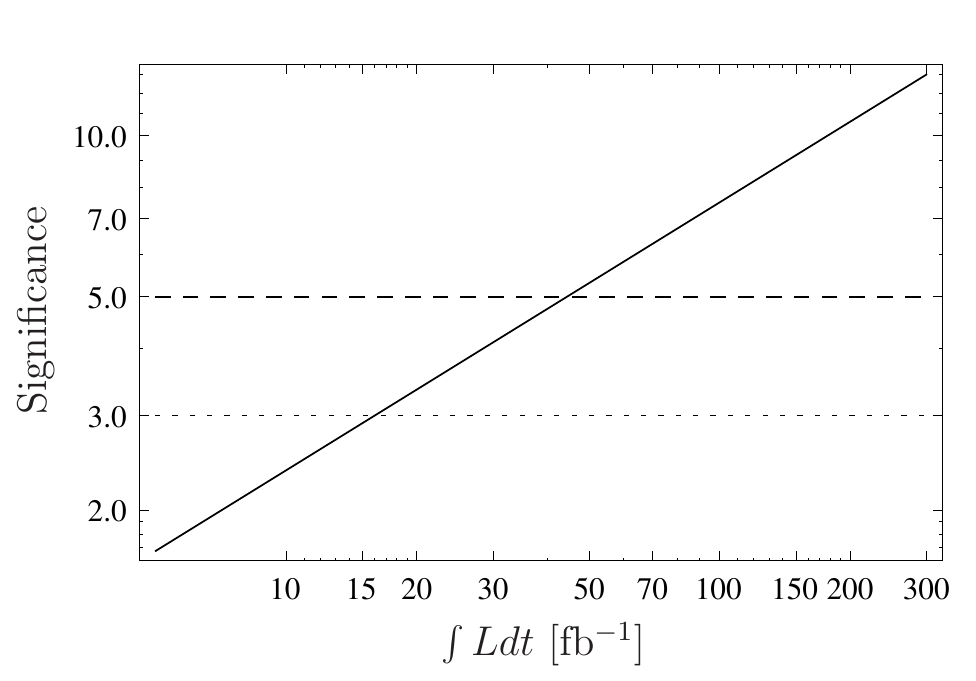}
\caption{Evolution of the significance level with growing integrated luminosity.
 The discussed significance at 100~fb$^{-1}$ fixes the curve. The borders for 3
 and 5$\sigma$ are shown as dotted and dashed lines.}
\label{fig:significanceEvolution}
\end{figure}
As can be seen from \FIG~\ref{fig:significanceEvolution},
with a luminosity of about 45 fb$^{-1}$ one crosses the
$5\sigma$ level. These numbers have been obtained using tree-level
calculations and it is known that higher-order corrections are
important in this context, see \EG \cite{Fuks:2007gk} and references
therein. In case of nonsupersymmetric models one obtains
K-factors of about 1.2-1.4 depending on the details of the models
\cite{Fuks:2007gk,Accomando:2010fz}. However, as it is not obvious
 how corrections due to supersymmetric particles will change this
 or how they affect the background reactions, we will stick to
tree-level calculations here.

\begin{figure}[htbp]
\centering
\begin{minipage}[b]{8.1cm}
\includegraphics[width=6.8cm]{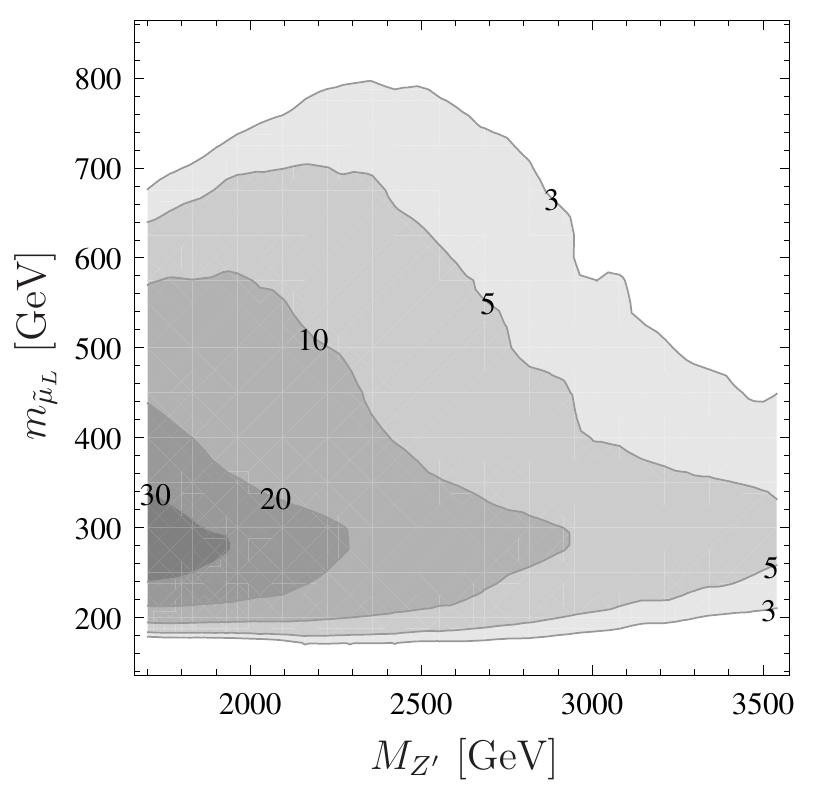}
\end{minipage}
\begin{minipage}[b]{8.1cm}
\includegraphics[width=6.8cm]{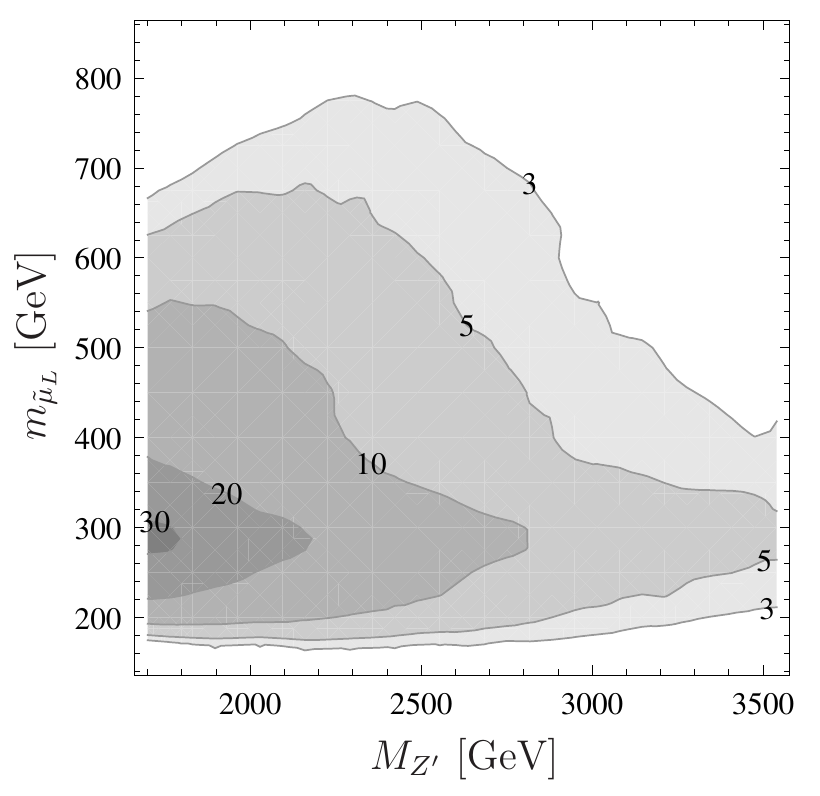}
\end{minipage}
\begin{minipage}[b]{8.1cm}
\includegraphics[width=6.8cm]{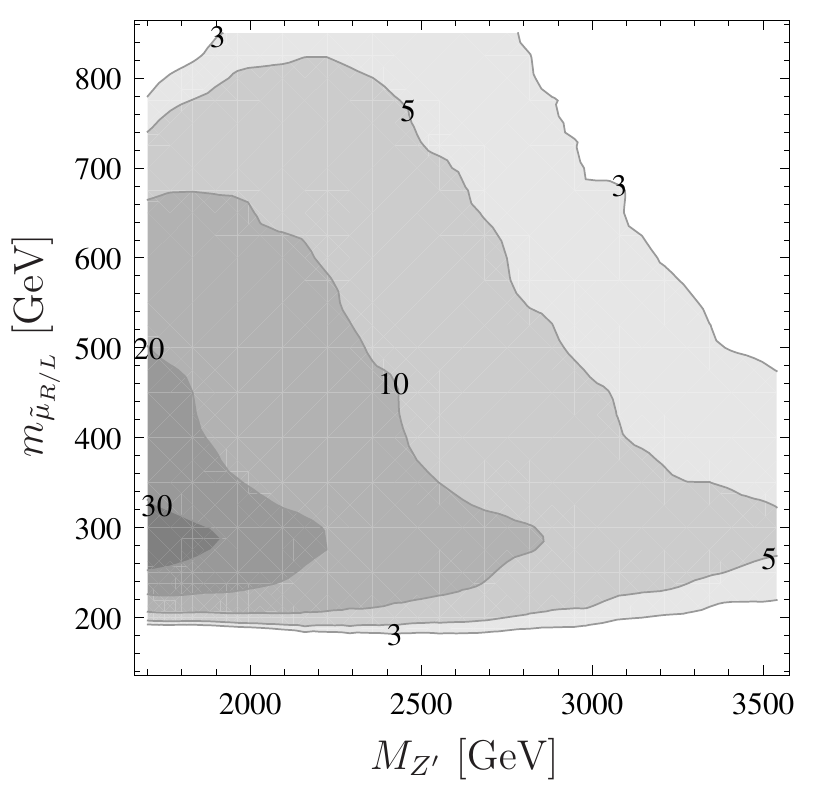}
\end{minipage}
\begin{minipage}[b]{8.1cm}
\includegraphics[width=6.8cm]{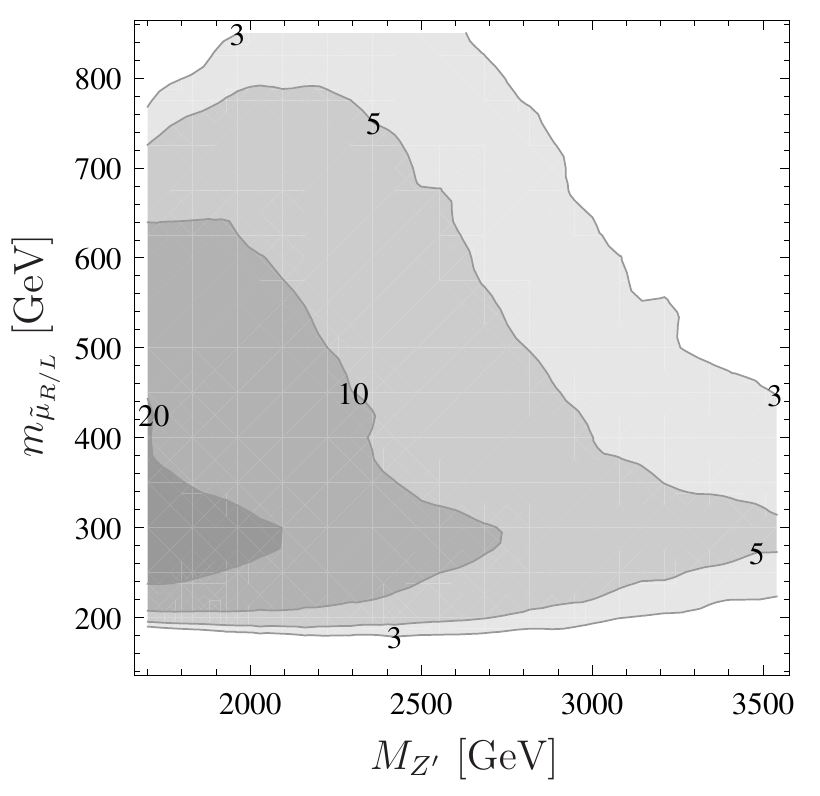}
\end{minipage}
\begin{minipage}[b]{8.1cm}
\includegraphics[width=6.8cm]{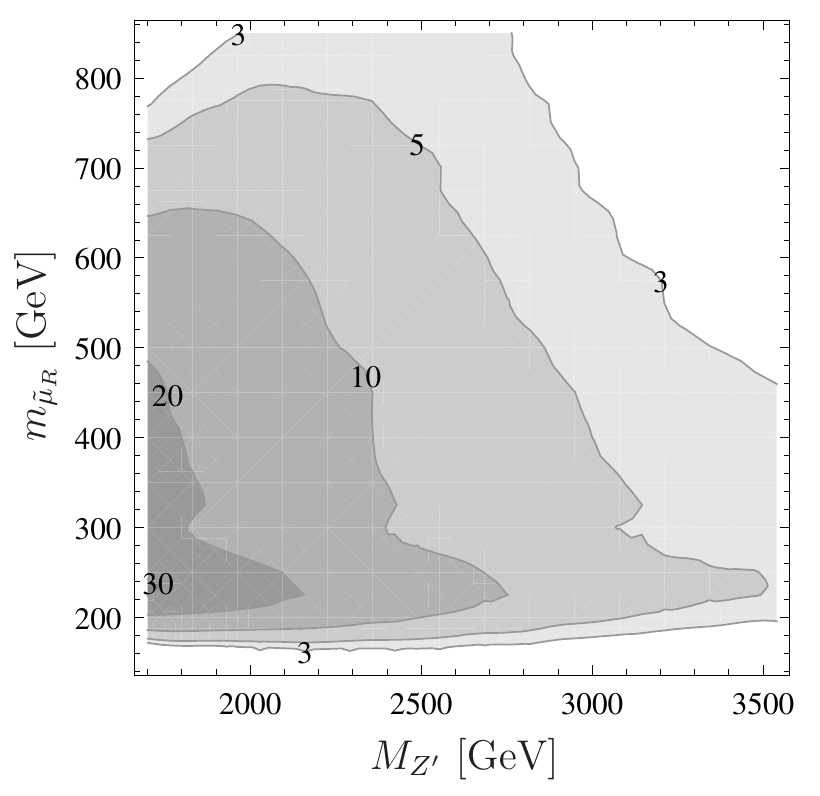}
\end{minipage}
\begin{minipage}[b]{8.1cm}
\includegraphics[width=6.8cm]{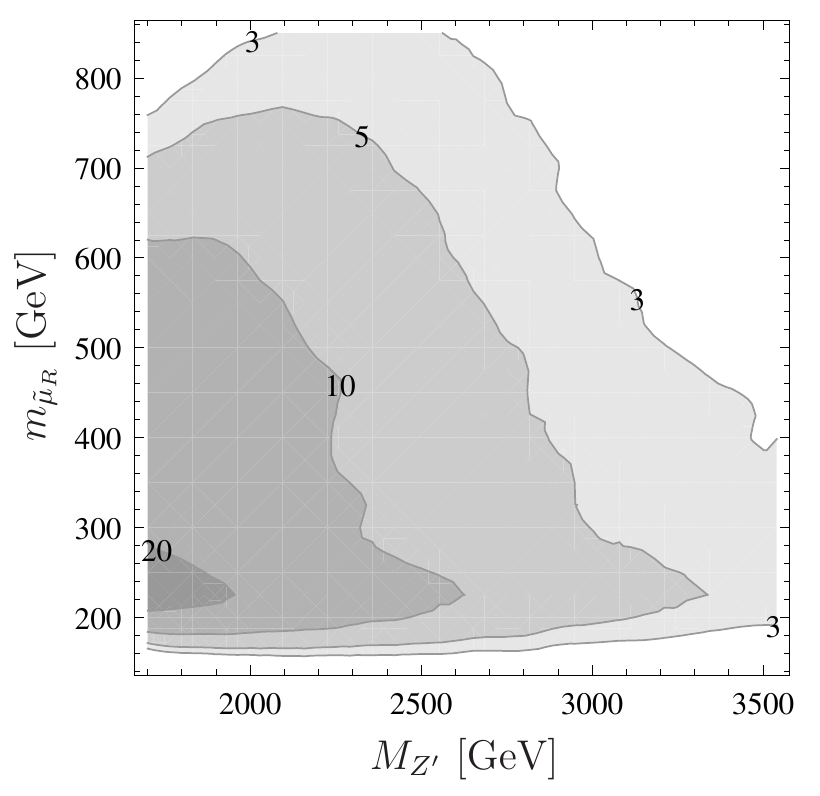}
\end{minipage}
\caption{Significance level of the smuon discovery as a function of the
 respective lightest $\tilde \mu$ mass and $M_{Z'}$ for
 ${\cal L}=100$~fb$^{-1}$ and $m_{{\tilde \chi}^0_1}=140$~GeV.
 Left column: looser cuts
 [$p_T(\emiss) > 200$~GeV, $M_{\mu \mu} > 200$~GeV, $M_T > 800$~GeV].
 Right column: tighter cuts 
[$p_T(\emiss) > 250$~GeV, $M_{\mu \mu} > 300$~GeV]. 
The smuon mass relations are (first row) 
$m_{\tilde \mu_R} = 1.2\, m_{\tilde \mu_L}$, 
(second row) $m_{\tilde \mu_{R}} = m_{\tilde \mu_L}$ and (last row) 
$m_{\tilde \mu_L} = 1.2\, m_{\tilde \mu_R}$.}
\label{fig:100fb}
\end{figure}

\begin{figure}[htbp]
\centering
\begin{minipage}[b]{8.1cm}
\includegraphics[width=7cm]{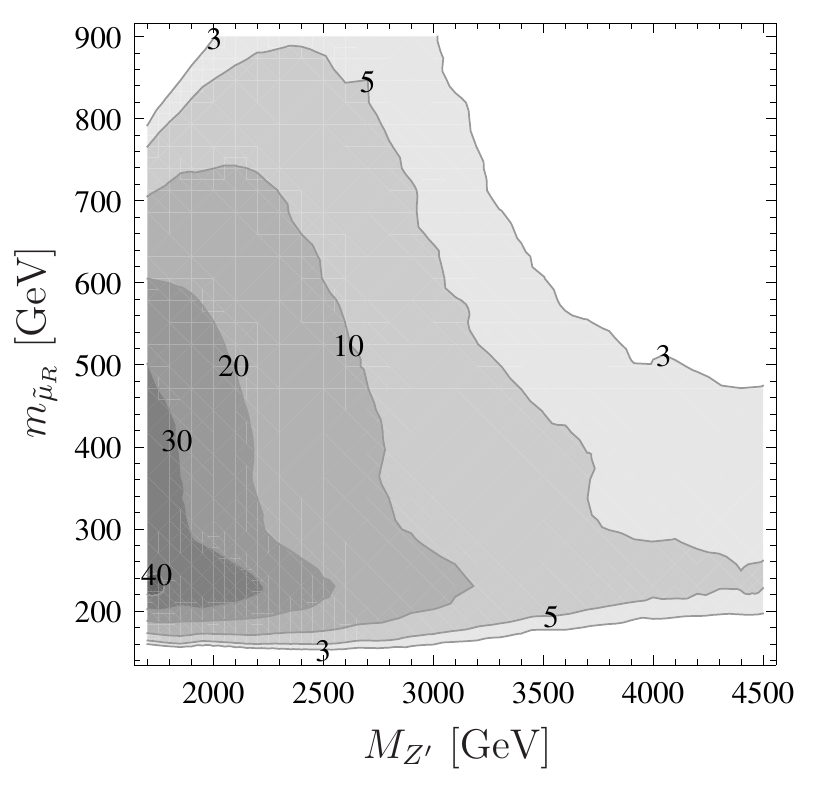}
\end{minipage}
\begin{minipage}[b]{8.1cm}
\includegraphics[width=7cm]{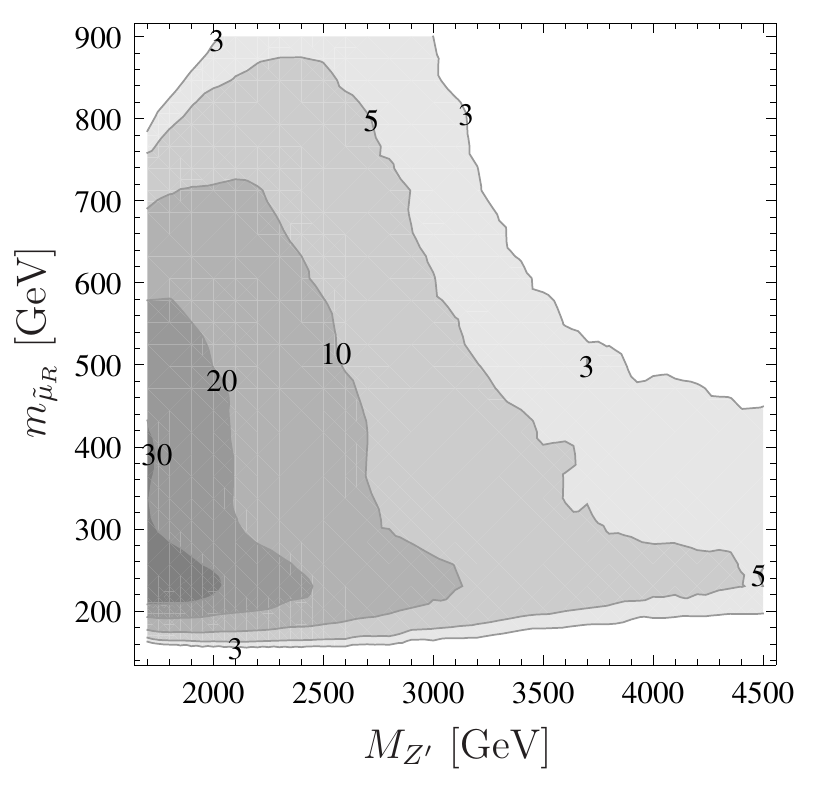}
\end{minipage}
\caption{Significance level of the smuon discovery  for
 ${\cal L}=300$~fb$^{-1}$, $m_{{\tilde \chi}^0_1}=140$~GeV  
 $m_{\tilde \mu_L}=1.2\, m_{\tilde \mu_R}$. Left: looser
 cuts [$p_T(\emiss) > 200$~GeV, $M_{\mu \mu} > 200$~GeV, $M_T > 800$~GeV].
 Right: tighter cuts [$p_T(\emiss) > 250$~GeV, 
 $M_{\mu \mu} > 300$~GeV]. The mass of the lightest supersymmetric particle is $140$~GeV.}
\label{fig:300fb}
\end{figure}

\begin{figure}[htbp]
\centering
\begin{minipage}[b]{8.1cm}
\includegraphics[width=7cm]{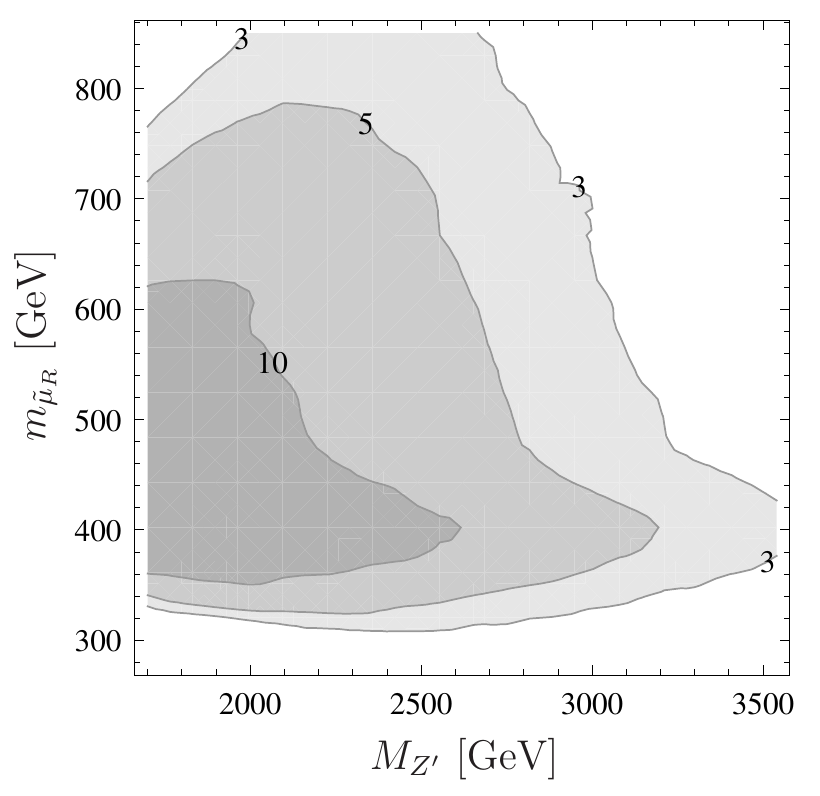}
\end{minipage}
\begin{minipage}[b]{8.1cm}
\includegraphics[width=7cm]{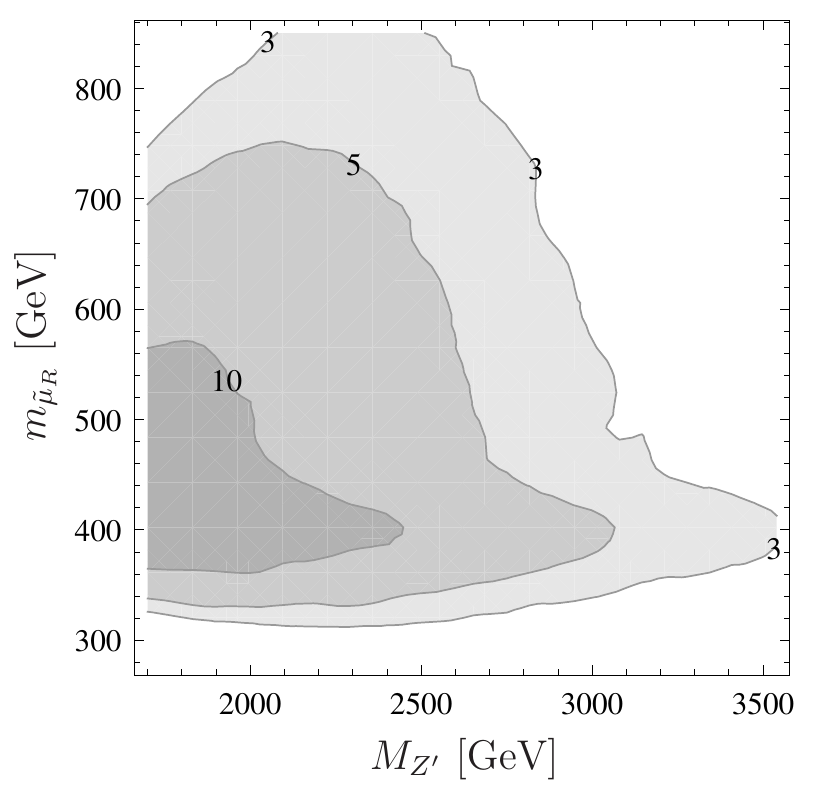}
\end{minipage}
\caption{Significance level of the smuon production for
 ${\cal L}=100$~fb$^{-1}$, $m_{{\tilde \chi}^0_1}=280$~GeV  
 $m_{\tilde \mu_L}=1.2\, m_{\tilde \mu_R}$. Left: looser
 cuts [$p_T(\emiss) > 200$~GeV, $M_{\mu \mu} > 200$~GeV, $M_T > 800$~GeV].
 Right: tighter cuts [$p_T(\emiss) > 250$~GeV, 
 $M_{\mu \mu} > 300$~GeV].}
\label{fig:100fb_280}
\end{figure}

The results depend so far mainly on the following quantities:
$M_{Z'}$, $m_{\tilde l}$, $m_{\tilde \chi^0_i}$ and
$m_{\tilde \chi^+_j}$
and on the nature of the neutralinos and charginos. 
We assume in the following that the lightest
two neutralinos and the lighter chargino are mainly the MSSM gauginos
as in our study points.
However,  we will depart to some
extent from the GUT assumptions: we will fix the masses of squarks
and gluinos to the values they take in benchmark
 point BLVI in \TAB~\ref{tab:parameter_points}
but vary the slepton masses freely. As we are interested 
in relatively light sleptons with masses down to 200 GeV
we fix $m_{\tilde \chi^0_1} = 140$~GeV and 
$m_{\tilde \chi^0_2} = m_{\tilde \chi^+_1} = 2\, m_{\tilde \chi^0_1}$.
We have performed a scan over slepton masses and $M_{Z'}$ fixing
the ratio of the masses for $R$ sleptons to
 $L$ sleptons
to 1.2, 1, and 1/1.2. The different ratios can  in principle
be obtained by
varying $\tan\beta'$, as can also be seen from the formulas
in Sec. \ref{subsec:sleptons}.\footnote{Note, however, that
$\tan\beta' < 1$ can only be obtained with non-universal 
bilepton mass parameters at the GUT scale.}
We have  also investigated the effect of tightening the previous
 cuts to 
$p_T(\emiss)> 250~\text{GeV}$ and $M_{\mu \mu}>300~\text{GeV}$. 
In \FIG~\ref{fig:100fb} we show our results for the significance
assuming an integrated luminosity of 100 fb$^{-1}$. We find a significant
dependence on the ratio of the smuon masses which is due to the fact
that we consider here only smuon decays into $\mu \tilde \chi^0_1$.
This is also the reason for the ``promontory'' for slepton masses up
to 300 GeV as there both $L$ and $R$ smuons can decay only into this
final state. 
The regions should even become somewhat larger if one includes also the
smuon decays into $\mu \tilde \chi^0_2$ and $\nu \tilde \chi^-_1$.
However, one has then to consider the background of squark and
gluino cascade decays that depend on the details of the parameter
point under study.

In \FIG~\ref{fig:300fb} we display the same for the case of
$m_{\tilde \mu_L} = 1.2\,m_{\tilde \mu_R}$ but taking a luminosity of 300 fb$^{-1}$.
As expected, the reach increases for both $m_{\tilde \mu}$ and
$M_{Z'}$. For the other two ratios of
slepton masses we find the same behavior.
In \FIG~\ref{fig:100fb_280} we show the changes for
increased neutralino and chargino masses
$m_{\tilde \chi^0_1} = 280$~GeV and 
${\tilde \chi^0_2}={\tilde \chi^+_1}=475~$GeV, which are the values obtained in the constrained model at BLVI.
As expected, the 5$\sigma$ significance is restricted to larger
values of the smuon mass as the leptons are softer than in
the previous example.

\section{Conclusion}
\label{sec:conclusions}

We have studied a supersymmetric model where the gauge
group is extended by an additional \UBL factor
resulting in a $Z'$ with a mass in the TeV range.
Such models can emerge as effective models from heterotic
string models. An important feature of this class of models
is gauge kinetic mixing. Here we focused on its impact for
the phenomenology of the extra vector boson. We have first shown
that bounds on its mass due to collider searches get significantly
reduced once the gauge kinetic mixing is taken into account:
LEP bounds by about 400 GeV and LHC bounds by about 200 GeV. 
Moreover, this implies that now LHC bounds are  more important
than the ones originating from LEP data. Second we have shown 
that these bounds get further reduced if the $Z'$ can decay into
supersymmetric particles. In our model the most important decays are
into sleptons, neutralinos and charginos. 

Moreover, we have discussed the reach of LHC with $\sqrt{s}=14$~TeV
for the discovery of smuons, which is an important possibility
should squark or gluino cascades in sufficient quantities be
 inaccessible.
For an integrated luminosity of 100~fb$^{-1}$ (300~fb$^{-1}$)
we have found that sleptons with masses of up to 800 GeV (900 GeV)
can be discovered this way, provided the $Z'$ is lighter than about
2.8 TeV (3.1 TeV). 
This result depends only mildly on the nature of the
smuons provided that their decays into muons
are not suppressed. 
Apart from detector effects, similar results hold
 for selectrons.
Staus, on the other hand, require a detailed study
 and we expect
a reduced reach at the LHC due to the hadronic decays of the 
resulting $\tau$ leptons.
 
\section*{Acknowledgments}

We thank Martin Hirsch, JoAnne L.\ Hewett, and Thomas G.~\ Rizzo 
for interesting discussion. We thank R.~Str\"ohmer for
clarifying remarks on the calculation of the significance.
W.P.~thanks the IFIC for hospitality during an extended stay
and the Alexander von Humboldt foundation for financial support.
This work has been supported by the
German Ministry of Education and Research (BMBF) under Contract 
No.\ 05H09WWEF.

\begin{appendix}
\section{$Z'$ couplings}
\label{sec:couplings}
In this section we collect the formulas for the couplings of $Z'$
to the fermions and scalars in this model.

\subsection{Couplings to fermions}
\label{sec:fcouplings}
The couplings given below follow from the terms in
 the Lagrangian
\begin{equation}
\bar{f_i} \gamma^\mu (c_{Lf,ij} P_L + c_{Rf,ij} P_R) f_j Z'_\mu \, .
\end{equation}
\begin{itemize}
\item Charged leptons: $Z' - \bar e_i - e_j$
\begin{align} 
 c_{Le,ij} =& \frac{1}{2} \delta_{i j} \Big(\Big(g_1\sin\Theta_W   - g_2 \cos\Theta_W  \Big)\sin{\Theta'}_W   + \Big(\gmix + \gBL{}\Big)\cos{\Theta'}_W  \Big)\, ,\\ 
 c_{Re,ij} =&  \,\frac{1}{2} \delta_{i j} \Big( 2 g_1 \sin\Theta_W  \sin{\Theta'}_W   + \Big(2 \gmix  + \gBL{}\Big)\cos{\Theta'}_W  \Big)\,.\end{align}

\item  Neutrinos: $Z' - \nu_i - \nu_j $
\begin{align} 
 c_{L\nu ,ij} =& \frac{1}{2} \Big(\Big( \Big(g_1 \sin\Theta_W   + g_2 \cos\Theta_W  \Big)\sin{\Theta'}_W   + \Big(\gmix + \gBL{}\Big)\cos{\Theta'}_W  \Big)\sum_{a=1}^{3}U^{V,*}_{j a} U_{{i a}}^{V}  \nonumber \\ 
 &- \gBL{} \cos{\Theta'}_W  \sum_{a=1}^{3}U^{V,*}_{j 3 + a} U_{{i 3 + a}}^{V}  \Big) \, ,\\ 
 c_{R\nu, ij} =& \,-\frac{1}{2} \Big(\Big( \Big(g_1\sin\Theta_W   + g_2 \cos\Theta_W  \Big)\sin{\Theta'}_W   + \Big(\gmix + \gBL{}\Big)\cos{\Theta'}_W  \Big)\sum_{a=1}^{3}U^{V,*}_{i a} U_{{j a}}^{V}  \nonumber \\ 
 &- \gBL{} \cos{\Theta'}_W  \sum_{a=1}^{3}U^{V,*}_{i 3 + a} U_{{j 3 + a}}^{V}  \Big)\,.\end{align} 
$U^V_{kl}$ is the unitary $6\times 6$ matrix that diagonalizes the
 neutrino mass matrix.

\item Up-type quarks: $Z' - \bar u_{i\alpha} - u_{j \beta} $
\begin{align} 
c_{Lu,ij} =& -\frac{1}{6} \delta_{\alpha \beta} \delta_{i j} \Big(\Big(-3 g_2 \cos\Theta_W   + g_1 \sin\Theta_W  \Big)\sin{\Theta'}_W   + \Big(\gmix + \gBL{}\Big)\cos{\Theta'}_W  \Big)\, ,\\ 
 c_{Ru,ij} =&  \,-\frac{1}{6} \delta_{\alpha \beta} \delta_{i j} \Big( 4 g_1 \sin\Theta_W  \sin{\Theta'}_W   + \Big(4 \gmix  + \gBL{}\Big)\cos{\Theta'}_W  \Big)\,.\end{align} 

\item Down-type quarks: $Z' - \bar d_{i\alpha} -  d_{j\beta} $
\begin{align} 
c_{Ld,ij} =& -\frac{1}{6} \delta_{\alpha \beta} \delta_{i j} \Big( \Big(3 g_2 \cos\Theta_W   + g_1 \sin\Theta_W  \Big)\sin{\Theta'}_W   + \Big(\gmix + \gBL{}\Big)\cos{\Theta'}_W  \Big)\, ,\\ 
 c_{Rd,ij} =& \,\frac{1}{6} \delta_{\alpha \beta} \delta_{i j} \Big(2 g_1 \sin\Theta_W  \sin{\Theta'}_W   + \Big(2 \gmix  - \gBL{}\Big)\cos{\Theta'}_W  \Big)\,.\end{align} 

\item Neutralinos: $Z' - \tilde \chi^0_i - \tilde \chi^0_j $
\begin{align} 
c_{L\tilde \chi^0,ij} =& \frac{1}{2} \Big(N^*_{j 3} \Big( \Big(g_1 \sin\Theta_W   + g_2 \cos\Theta_W  \Big)\sin{\Theta'}_W   + \gmix \cos{\Theta'}_W  \Big)N_{{i 3}} \nonumber \\ 
 &-N^*_{j 4} \Big(g_1 \sin\Theta_W  \sin{\Theta'}_W   + g_2 \cos\Theta_W  \sin{\Theta'}_W   + \gmix \cos{\Theta'}_W  \Big)N_{{i 4}} \nonumber \\ 
 &+2 \gBL{} \cos{\Theta'}_W   \Big(N^*_{j 6} N_{{i 6}}  - N^*_{j 7} N_{{i 7}} \Big)\Big)\, ,\\ 
 c_{R\tilde \chi^0,ij} =&  \,-\frac{1}{2} \Big(N^*_{i 3} \Big( \Big(g_1 \sin\Theta_W   + g_2 \cos\Theta_W  \Big)\sin{\Theta'}_W   + \gmix \cos{\Theta'}_W  \Big)N_{{j 3}} \nonumber \\ 
 &-N^*_{i 4} \Big(g_1 \sin\Theta_W  \sin{\Theta'}_W   + g_2 \cos\Theta_W  \sin{\Theta'}_W   + \gmix \cos{\Theta'}_W  \Big)N_{{j 4}} \nonumber \\ 
 &+2 \gBL{} \cos{\Theta'}_W  \Big(N^*_{i 6} N_{{j 6}}  - N^*_{i 7} N_{{j 7}} \Big)\Big)\,.\end{align} 
$N_{kl}$ is the unitary $7 \times 7$ matrix that diagonalizes the
 neutralino mass matrix.

\item Charginos: $Z' - \tilde \chi^+_i - \tilde \chi^-_j $
\begin{align} 
c_{L\tilde \chi^\pm,ij} =&\, - \frac{1}{2} \Big(2 g_2 U^*_{j 1} \cos\Theta_W  \sin{\Theta'}_W  U_{{i 1}} \nonumber \\ 
 &-U^*_{j 2} \Big(\Big( g_1 \sin\Theta_W   - g_2 \cos\Theta_W  \Big)\sin{\Theta'}_W   + \gmix \cos{\Theta'}_W  \Big)U_{{i 2}} \Big)\, ,\\ 
c_{R\tilde \chi^\pm,ij}  =&  \,-\frac{1}{2} \Big(2 g_2 V^*_{i 1} \cos\Theta_W  \sin{\Theta'}_W  V_{{j 1}} \nonumber \\ 
 &-V^*_{i 2} \Big(\Big( g_1 \sin\Theta_W   - g_2 \cos\Theta_W  \Big)\sin{\Theta'}_W   + \gmix \cos{\Theta'}_W  \Big)V_{{j 2}} \Big)\,.\end{align}
$U_{kl}$ and $V_{kl}$ are the unitary $2 \times 2$ matrices needed to
 diagonalize chargino mass matrix.
\end{itemize}

\subsection{Couplings to scalars}
The couplings given below follow from the terms in
 the Lagrangian
\begin{equation}
c_{s,ij} \tilde s_i  \tilde s_j^* 
\left(p_{s_i}^{\mu}- p_{s^*_j}^{\mu} \right) Z'_\mu\, ,
\end{equation}
where $p_{s_i}$ and $p_{s^*_j}$ are the four-momenta of the scalars.
In the following, $Z^p_{kl}$ denote the matrices needed
 to diagonalize the respective underlying mass matrix of the particles $p$.
\begin{itemize}
\item Charged sleptons: $Z' - \tilde e_i - \tilde e_j^* $
\begin{align}
c_{e,ij} =&\frac{1}{2} \Big(\Big(\Big( g_1 \sin\Theta_W   - g_2 \cos\Theta_W  \Big)\sin{\Theta'}_W   + \Big(\gmix + \gBL{}\Big)\cos{\Theta'}_W  \Big)\sum_{a=1}^{3}Z^{E,*}_{i a} Z_{{j a}}^{E}  \nonumber \\ 
 &+\Big( 2 g_1 \sin\Theta_W  \sin{\Theta'}_W   + \Big(2 \gmix  + \gBL{}\Big)\cos{\Theta'}_W  \Big)\sum_{a=1}^{3}Z^{E,*}_{i 3 + a} Z_{{j 3 + a}}^{E}  \Big)\,.\end{align} 

\item  Sneutrinos: $Z' - \tilde \nu^P_i - \tilde \nu_j^{S*} $
\begin{align} 
c_{\nu, ij} =&\frac{i}{2} \Big(- \Big( \Big(g_1 \sin\Theta_W   + g_2 \cos\Theta_W  \Big)\sin{\Theta'}_W   + \Big(\gmix + \gBL{}\Big)\cos{\Theta'}_W  \Big)\sum_{a=1}^{3}Z^{P,*}_{i a} Z^{S,*}_{j a}  \nonumber \\ 
 &- \gBL{} \cos{\Theta'}_W \sum_{a=1}^{3}Z^{P,*}_{i 3 + a} Z^{S,*}_{j 3 + a}  \Big)\,.\end{align}  

\item Up-type squarks: $Z' - \tilde u_{i \alpha} - \tilde u_{j \beta}^* $
\begin{align} 
c_{q_u,ij} =&-\frac{1}{6} \delta_{\alpha \beta} \Big(\Big(\Big(-3 g_2 \cos\Theta_W   + g_1 \sin\Theta_W  \Big)\sin{\Theta'}_W   + \Big(\gmix + \gBL{} \Big)\cos{\Theta'}_W  \Big)\sum_{a=1}^{3}Z^{U,*}_{i a} Z_{{j a}}^{U}  \nonumber \\ 
 &+\Big( 4 g_1 \sin\Theta_W  \sin{\Theta'}_W   + \Big(4 \gmix  + \gBL{}\Big)\cos{\Theta'}_W  \Big)\sum_{a=1}^{3}Z^{U,*}_{i 3 + a} Z_{{j 3 + a}}^{U}  \Big)\,.\end{align}

\item Down-type squarks: $Z' - \tilde d_{i \alpha} - \tilde d_{j \beta}^* $
\begin{align} 
c_{q_d,ij} =&-\frac{1}{6} \delta_{\alpha \beta} \Big(\Big( \Big(3 g_2 \cos\Theta_W   + g_1\sin\Theta_W  \Big)\sin{\Theta'}_W   + \Big(\gmix + \gBL{}\Big)\cos{\Theta'}_W  \Big)\sum_{a=1}^{3}Z^{D,*}_{i a} Z_{{j a}}^{D}  \nonumber \\ 
 &+\Big(-2 g_1\sin\Theta_W  \sin{\Theta'}_W   + \Big(-2 \gmix  + \gBL{}\Big)\cos{\Theta'}_W  \Big)\sum_{a=1}^{3}Z^{D,*}_{i 3 + a} Z_{{j 3 + a}}^{D}  \Big)\,.\end{align} 
 
\item Charged Higgs: $Z' - H^-_i - H^+_j$
\begin{align} 
c_{H^\pm ,ij} =&\frac{1}{2} \delta_{i j} \Big(\Big( g_1 \sin\Theta_W   - g_2 \cos
\Theta_W  \Big)\sin{\Theta'}_W   + \gmix \cos{\Theta'}_W  \Big)\,.\end{align} 

\item $CP$-odd and $CP$-even Higgs: $Z' - A^0_i - h_j$
\begin{align} 
c_{Ah,ij} =&\frac{i}{2} \Big(-\Big(\Big(g_1 \sin\Theta_W   + g_2 \cos\Theta_W
\Big)\sin{\Theta'}_W   + \gmix \cos{\Theta'}_W  \Big)Z_{{i 1}}^{A}
Z_{{j 1}}^{H} \nonumber \\ 
 &+\Big( \Big(g_1 \sin\Theta_W   + g_2 \cos\Theta_W
\Big)\sin{\Theta'}_W   + \gmix \cos{\Theta'}_W  \Big)Z_{{i 2}}^{A}
Z_{{j 2}}^{H} \nonumber \\ 
 &-2 \gBL{} \cos{\Theta'}_W   \Big(Z_{{i 3}}^{A} Z_{{j 3}}^{H}  - Z_{{i 4}}^{A}
Z_{{j 4}}^{H} \Big)\Big)\,.
\end{align}
\end{itemize}

\subsection{Coupling to vector bosons}
\label{sec:vcouplings}
The only three-vector-boson vertex containing a $Z'$ is the
 coupling $Z'_\mu - W^+_\rho - W^-_\sigma$. It is parametrized as follows:
\begin{equation}
c_{V_iV_j} \Big(g_{\rho \mu} \Big(-
p^{{Z'}_{{\mu}}}_{\sigma}  + p^{W^+_{{\rho}}}_{\sigma}\Big) + g_{\rho
\sigma} \Big(- p^{W^+_{{\rho}}}_{\mu}  + p^{W^-_{{\sigma}}}_{\mu}\Big) +
g_{\sigma \mu} \Big(- p^{W^-_{{\sigma}}}_{\rho}  +
p^{{Z'}_{{\mu}}}_{\rho}\Big)\Big) Z'^\mu V_i^\rho V_j^\sigma\, ,
\end{equation}
with
\begin{align}c_{WW} =  g_2 \cos\Theta_W  \sin{\Theta'}_W\, .\end{align}

\subsection{Coupling to one vector boson and one scalar}
\label{sec:vscouplings}
The vertices are parametrized as follows:
\begin{equation}
c_{Vs,i} s_i g_{\sigma \mu} Z'^\mu V^\sigma\, .
\end{equation}
\begin{itemize}
\item $Z$ and Higgs: $Z'_\mu - Z_\sigma - h_i$
\begin{align}
c_{Zh,i}&=\frac{1}{2} \Big(- v_d \Big( g_1 \gmix \cos{\Theta'}_{W }^{2} \sin
\Theta_W  +g_{2}^{2} \cos\Theta_{W }^{2} \cos{\Theta'}_W
\sin{\Theta'}_W  \nonumber \\ 
 &+\cos{\Theta'}_W  \Big(g_{1}^{2} \sin\Theta_{W }^{2}  - \gmix^{2}
\Big)\sin{\Theta'}_W  -g_1 \gmix \sin\Theta_W  \sin{\Theta'}_{W }^{2}
\nonumber \\ 
 &+g_2 \cos\Theta_W  \Big(g_1 \sin\Theta_W  \sin2 {\Theta'}_W    + \gmix \cos{\Theta'}_{W }^{2}  - \gmix \sin{\Theta'}_{W }^{2}
\Big)\Big)Z_{{i 1}}^{H} \nonumber \\ 
 &- v_u \Big( g_1 \gmix \cos{\Theta'}_{W }^{2} \sin\Theta_W
+g_{2}^{2} \cos\Theta_{W }^{2} \cos{\Theta'}_W  \sin{\Theta'}_W
\nonumber \\ 
 &+\cos{\Theta'}_W  \Big(g_{1}^{2} \sin\Theta_{W }^{2}  - \gmix^{2}
\Big)\sin{\Theta'}_W  -g_1 \gmix \sin\Theta_W  \sin{\Theta'}_{W }^{2}
\nonumber \\ 
 &+g_2 \cos\Theta_W  \Big(g_1 \sin\Theta_W  \sin2 {\Theta'}_W    + \gmix \cos{\Theta'}_{W }^{2}  - \gmix \sin{\Theta'}_{W }^{2}
\Big)\Big)Z_{{i 2}}^{H} \nonumber \\ 
 &+2 \gBL{} \sin2
{\Theta'}_W  \Big(v_{\eta}
Z_{{i 3}}^{H}  + v_{\bar \eta} Z_{{i 4}}^{H} \Big)\Big)\,.
\end{align}
\end{itemize}

\end{appendix}


\end{document}